\documentclass{svjour3} 
\usepackage{amsmath}

\newcommand{\toolname}{\textsc{CL4SE}\xspace}
\usepackage{wasysym}
\DeclareUnicodeCharacter{2642}{\male}

\usepackage{tcolorbox}
\tcbuselibrary{breakable}
\usepackage{pifont}
\usepackage{enumerate}
\usepackage{ragged2e}
\usepackage{stfloats}
\usepackage[vlined, ruled]{algorithm2e}
\usepackage{algorithmic}
\usepackage{graphicx}
\usepackage{textcomp}
\usepackage{multirow}
\usepackage{subfigure}
\usepackage{diagbox}
\usepackage{listings}
\usepackage{float}
\usepackage{url}
\usepackage{hyperref}
\usepackage{amsmath}

\definecolor{deepblue}{rgb}{0,0,0.5}
\definecolor{deepgreen}{rgb}{0,0.5,0}
\definecolor{deepred}{rgb}{0.6,0,0}
\definecolor{darkorange}{RGB}{255,140,0}
\definecolor{lightgray}{rgb}{0.93,0.93,0.93}
\definecolor{deepgray}{rgb}{0.25,0.25,0.25}

\hypersetup{
    colorlinks=true,
    linkcolor=blue,
    filecolor=blue,      
    urlcolor=blue,
    citecolor=blue,
}

\lstdefinelanguage{Java}{
	basicstyle=\small\ttfamily,
	numberstyle=\color{deepgray},
	stepnumber=1,
	numbersep=8pt,
	showstringspaces=false,
	breaklines=true,
	frame=lines,
	backgroundcolor=\color{lightgray},
	commentstyle=\color{deepgreen},
	keywordstyle=\color{deepblue},
	stringstyle=\color{deepred},
	tabsize=4,
	captionpos=b,
	morekeywords={public, class, void, int, if, else, for, while, return, true, false},
	emph={String, System},
	emphstyle=\color{darkorange},
	alsoletter={.,;:[]()},
}
\usepackage{multirow}
\usepackage[utf8]{inputenc}
\usepackage{graphicx}
\usepackage{textcomp}
\usepackage{color,xcolor}
\usepackage{url}
\usepackage{graphicx}
\usepackage{subfigure}
\usepackage{stmaryrd}
\usepackage{verbatimbox}
\usepackage{enumerate}
\usepackage[shortlabels]{enumitem}
\usepackage{bm}
\usepackage[justification=centering]{caption}
\usepackage{makecell}
\usepackage{booktabs}

\newcommand{\finding}[2]{
\begin{center}
\begin{tcolorbox}[leftrule=0mm,toprule=0mm,bottomrule=0mm,rightrule=0mm,left=1pt,right=2pt,top=0pt,bottom=0pt,breakable]
\textbf{Answer to RQ{#1}:}
{#2}
\end{tcolorbox}
\end{center}
}

\usepackage{multirow}

\usepackage{enumitem}
\usepackage{colortbl}
\usepackage{natbib}
\usepackage[normalem]{ulem}

\newcommand{\delete}[1]{}

\journalname{Empirical Software Engineering}
\date{Received: date / Accepted: date}
\title{\toolname{}: Benchmarking Context Learning on Software Engineering}

\author{Haichuan Hu \and 
Quanjun Zhang \and
Ye Shang \and
Guoqing Xie \and
Chunrong Fang \and
Zhenyu Chen  \and
Liang Xiao \and
}

\institute{ Haichuan Hu \at
            School of Computer Science and Engineering, Nanjing University of Science and Technology, China\\
              \email{huhaichuan2024@gmail.com}
            \and
            Quanjun Zhang \at
              School of Computer Science and Engineering, Nanjing University of Science and Technology, China \\
              \email{quanjunzhang@njust.edu.cn}
           \and
           Ye Shang \at
           State Key Laboratory for Novel Software Technology, Nanjing University, China\\
           \email{yeshang@smail.nju.edu.cn}
           \and
           Guoqing Xie \at
           State Key Laboratory for Novel Software Technology, Nanjing University, China\\
           \email{522025320181@smail.nju.edu.cn}
           \and
           Chunrong Fang \at
           State Key Laboratory for Novel Software Technology, Nanjing University, China\\
           \email{fangchunrong@nju.edu.cn}
           \and
           Zhenyu Chen \at
           State Key Laboratory for Novel Software Technology, Nanjing University, China\\
           \email{zychen@nju.edu.cn}
           \and
           Liang Xiao \at
           School of Computer Science and Engineering, Nanjing University of Science and Technology, China \\
           \email{xiaoliang@mail.njust.edu.cn}
}

\begin{document}

\maketitle

\abstract{
Context engineering has emerged as a pivotal paradigm for unlocking the potential of Large Language Models (LLMs) in Software Engineering (SE) tasks, enabling performance gains at test time without model fine-tuning. Despite its success, existing research lacks a systematic taxonomy of SE-specific context types and a dedicated benchmark to quantify the heterogeneous effects of different contexts across core SE workflows. To address this gap, we propose \toolname{} (Context Learning for Software Engineering), a comprehensive benchmark featuring a fine-grained taxonomy of four SE-oriented context types (interpretable examples, project-specific context, procedural decision-making context, and positive \& negative context), each mapped to a representative task (code generation, code summarization, code review, and patch correctness assessment). We construct high-quality datasets comprising over 13,000 samples from more than 30 open-source projects and evaluate five mainstream LLMs across nine metrics. Extensive experiments demonstrate that context learning yields an average performance improvement of 24.7\% across all tasks. Specifically, procedural context boosts code review performance by up to 33\% (Qwen3-Max), mixed positive-negative context improves patch assessment by 30\% (DeepSeek-V3), project-specific context increases code summarization BLEU by 14.78\% (GPT-Oss-120B), and interpretable examples enhance code generation PASS@1 by 5.72\% (DeepSeek-V3). Beyond these task-level gains, our results show that the effectiveness of context depends strongly on its alignment with task-specific cognitive demands, that more context does not necessarily lead to better performance, and that reasoning-intensive SE tasks benefit most from well-structured contextual support. \toolname{} establishes the first standardized evaluation framework for SE context learning, provides actionable empirical insights into task-specific context design, and releases a large-scale dataset to facilitate reproducible research in this domain.

}

\section{Introduction}
With the rapid advancement of Large Language Models (LLMs), context engineering~\citep{mei2025survey,zhang2025agentic} has emerged as a pivotal paradigm for unlocking their potential in downstream tasks. Unlike pre-training or fine-tuning, which encode knowledge via model weight updates, context engineering improves LLM performance at inference time through the construction of effective contextual inputs, without modifying the model parameters. In Software Engineering (SE), researchers have increasingly leveraged this paradigm together with models’ In-Context Learning (ICL) capabilities to improve performance across a wide range of development tasks~\citep{li2025large,gao2025sva,wu2025empirical,tang2024context,xu2024unilog,yun2024project,geng2024large,shao2026fix}. For example,~\citep{li2025large} trains a model-aware retriever to select more effective exemplars for code generation, while~\citep{shao2026fix} incorporates relevant repair patterns as context to guide models toward more effective vulnerability repair.

\begin{figure*}[htbp]
  \centering
  \includegraphics[width=0.7\linewidth]{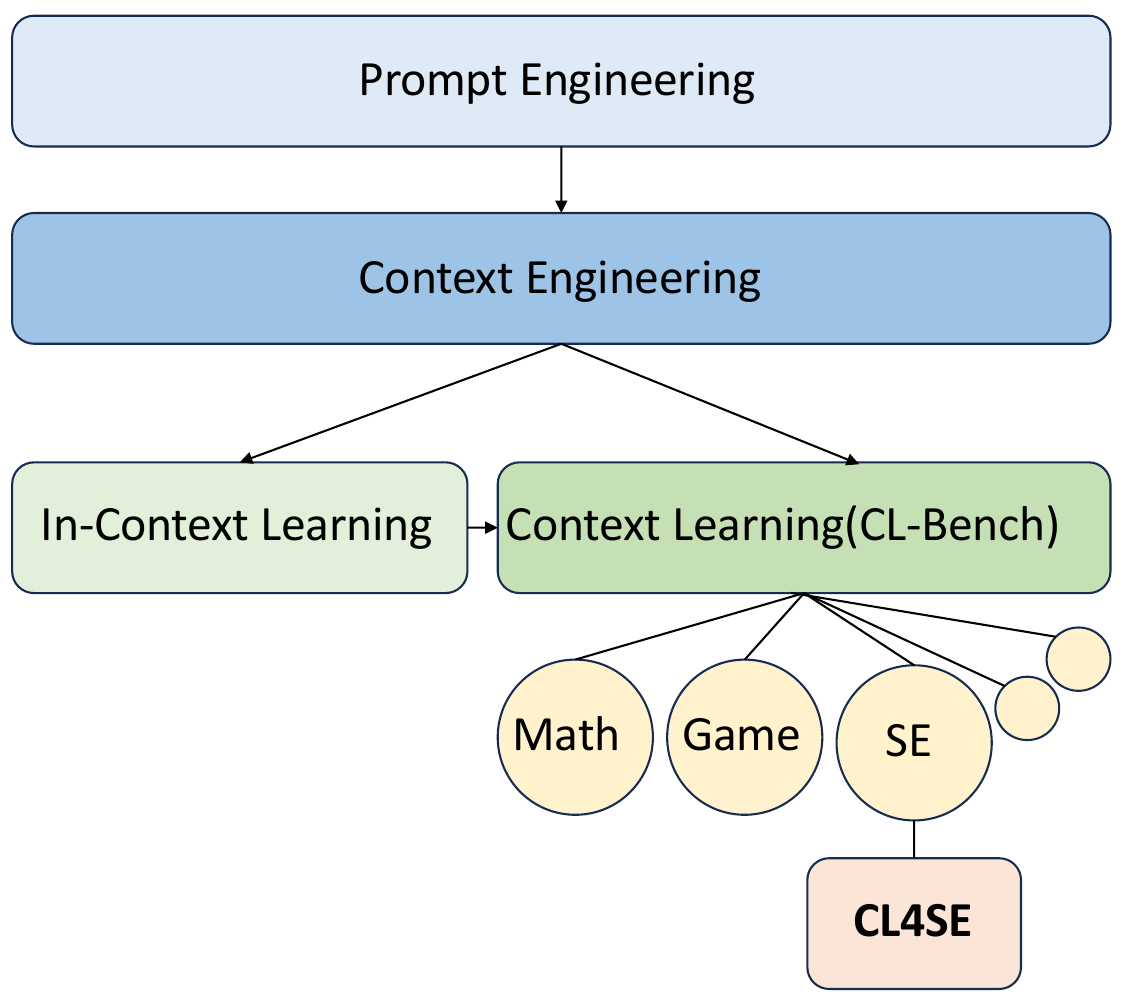}
  \caption{The relationship between \toolname{}, prompt engineering, context engineering, in-context learning and context learning.} 
  \label{fig:cl4se}
\end{figure*}

Despite the demonstrated effectiveness of ICL and context engineering in SE tasks, existing research still lacks a systematic understanding of how different context types affect different tasks, and of the mechanisms through which such contexts take effect. Early work~\citep{gao2023makes} explored how to construct in-context demonstrations more effectively, but did not establish a systematic taxonomy of SE-specific contexts or analyze how models learn from them. More recently, CL-Bench~\citep{dou2026cl} extended traditional ICL research by investigating the patterns and effectiveness of context learning for complex tasks, particularly in settings where prior knowledge is limited. Inspired by this line of work, we aim to fill the critical gap in context learning for software engineering by proposing \toolname{}, a dedicated benchmark tailored to SE tasks. Figure~\ref{fig:cl4se} illustrates the relationship between \toolname{} and related concepts, including prompt engineering, context engineering, in-context learning, and context learning. As an SE-oriented extension of CL-Bench, \toolname{} goes beyond generic context learning benchmarks for general NLP or broad reasoning tasks. Instead, it systematically characterizes heterogeneous SE-specific context types and examines how they guide LLMs’ in-context learning across core SE workflows.

To this end, we define a fine-grained taxonomy of four SE-oriented context types, namely interpretable examples, project-specific context, procedural decision-making context, and positive \& negative context, each aligned with a representative SE task: code generation, code summarization, code review, and patch correctness assessment. We further construct high-quality real-world datasets with over 13,000 samples collected from more than 30 open-source projects, and evaluate five mainstream LLMs using nine metrics. Extensive experiments show that each context type exhibits a distinct efficacy profile aligned with the cognitive demands of its target task. Interpretable reasoning context is most effective for complex code generation, project-specific context substantially improves stylistic alignment in code summarization, procedural decision context yields strong gains in code review, and dual positive-negative context improves the precision-recall balance in patch correctness assessment. Overall, the results confirm that context learning is not a one-size-fits-all strategy, with our benchmark revealing an average performance improvement of 24.7\% across tasks and gains of up to 33\% in decision-centric workflows.

In addition, our study leads to the following key findings:

\begin{itemize}[leftmargin=*]
    \item \textbf{The effectiveness of context depends on task-context alignment.} Context becomes most useful when its structure matches the cognitive demands of the target task, suggesting that context learning in SE should be designed in a task-aware rather than task-agnostic manner.

    \item \textbf{More context does not necessarily lead to better performance.} In many cases, such as code summarization, a small amount of well-targeted context is more effective than a large amount of redundant context. This suggests that the quality and relevance of context matter more than its volume.

    \item \textbf{Context learning is especially valuable for reasoning-intensive tasks.} The strongest benefits arise in tasks involving judgment, comparison, and multi-step reasoning, indicating that context can shape not only what models output, but also how they reason.

    \item \textbf{Effective context teaches both solutions and reasoning processes.} The most informative contexts do not merely provide correct outputs; they also expose reasoning paths, decision procedures, and contrasts between valid and invalid cases, enabling models to learn both what to do and how to do it.

    \item \textbf{Context learning should be recognized as a core capability for SE intelligence.} Our findings suggest that context is not merely an auxiliary prompt component, but a central medium through which LLMs acquire and apply task-relevant knowledge at inference time. In SE settings, the ability to learn effectively from context may be as important as the model’s parametric knowledge itself.
\end{itemize}

Our work not only establishes a standardized evaluation framework for future research on SE-focused context learning, but also provides actionable empirical insights into how to construct effective contextual inputs, enabling researchers and practitioners to better unlock LLMs’ potential in real-world software development through principled context design rather than ad-hoc prompt crafting.

\section{Related Work}
\subsection{Prompt/Context Engineering}

\textbf{Prompt Engineering.} Traditional prompt engineering~\citep{giray2023prompt,marvin2023prompt,chen2025unleashing,liu2023pre,sahoo2024systematic} is the process of designing and refining input queries to elicit desired responses from LLMs. This paradigm aims to guide models to better leverage parameterized knowledge, rather than introducing new content that requires learning into the prompt. Subsequent works~\citep{wang2024advanced,syahputri2025unlocking,ronanki2025prompt,rodriguez2023prompts,liu2024logprompt,wang2024enhancing} have discussed the application of Prompt Engineering in the context of Software Engineering. However, with the emergence of the latest reasoning models (e.g., GPT-o1~\citep{hu2025can}), researchers~\citep{wang2024advanced} have found that earlier prompt techniques may no longer yield significant effects and may even lead to negative returns.

\textbf{Context Engineering.} Nowadays, a growing body of researchers is moving away from prompt engineering and towards context engineering~\citep{mei2025survey,zhang2025agentic,dai2025onepiece}. Unlike prompt engineering that prioritizes prompt structure and crafting tricks, context engineering shifts the focus to the full lifecycle of prompt content. It revolves around retrieving~\citep{singh2025agentic}, organizing, managing~\citep{xu2025mem,xu2025everything}, and optimizing task-relevant context sourced from heterogeneous repositories including private documents, databases, and knowledge bases. For example, in complex code repair tasks that require long contexts, researchers~\citep{xia2024agentless} use hierarchical positioning techniques to compress the context. Code repository analysis based on knowledge graphs~\citep{ouyang2024repograph} facilitates the rapid acquisition of context in various SE tasks. With the rise of agent technologies, context engineering has become increasingly important. Agents require continuous, multi-turn context understanding and management, which imposes higher requirements on the effectiveness and efficiency of context. Recent studies~\citep{liu2026intent,cheng2026contextual} show that in long multi-turn dialogues, early context errors can trigger a butterfly effect, where minor initial mistakes gradually propagate, accumulate, and eventually lead to severe degradation in reasoning accuracy and task performance. This further motivates the need for dedicated research into effective context engineering principles tailored for advanced LLMs and agentic workflows.

\textbf{In-Context Learning and Context Learning.}
ICL~\citep{dong2024survey,akyurek2022learning} serves as the fundamental paradigm that underpins both prompt engineering and context engineering, acting as the core mechanism through which LLMs generalize to unseen tasks without parameter updates. Traditional prompt engineering primarily leverages ICL by manually crafting a small set of exemplars within the prompt to guide the model's output~\citep{brown2020language}, treating context as a static input component. In contrast, context engineering elevates the role of ICL by systematically curating, structuring, and maintaining high-quality contextual information, thereby enabling more robust and scalable ICL performance in complex scenarios such as long-horizon code repair and multi-turn agent interactions. Building upon the advancements in context engineering, the concept of \textit{Context Learning (CL)}~\citep{dou2026cl} has emerged as a broader and more principled framework that transcends the limitations of traditional ICL. Rather than merely relying on ad-hoc exemplar selection as in ICL,CL emphasizes the ability of models to learn and apply new knowledge from scarce or complex data.

Accordingly, our work deepens and refines the study of context learning in the field of software engineering. We move beyond generic context learning benchmarks by constructing a specialized evaluation framework that captures the unique scarcity and complexity of data in real-world SE workflows. This refined focus enables us to provide empirical insights into the practical application and effectiveness of context engineering for advanced LLMs and agent systems in the SE domain.

\subsection{LLM-based Software Engineering}

The integration of LLMs has profoundly reshaped the landscape of SE. Modern SE encompasses a systematic lifecycle primarily including requirement and design~\citep{wang2024application}, development~\citep{mathews2024test,joel2024survey}, testing~\citep{chen2024chatunitest,nan2025test}, maintenance~\citep{patil2025advancing,bouzenia2024repairagent,hu2025tsapr}, and management~\citep{zhang2023survey,he2025llm,hu2025repair,zhang2026compass}. LLMs have demonstrated remarkable capabilities across SE lifecycle, particularly in software development and software maintenance, which heavily rely on code understanding, reasoning, and generation. In this paper, we focus on four representative SE tasks that span these critical phases: code generation, code summarization, code review, and patch correctness assessment.

During the software development phase, code generation and code summarization are two fundamental tasks. For code generation~\citep{crupi2025effectiveness,joel2024survey}, recent advancements have shifted from direct sequence-to-sequence translation to more sophisticated requirement-guided and execution-guided paradigms~\citep{jiang2024self,dong2024self}. Researchers increasingly rely on prompt engineering and in-context learning to help LLMs understand complex intents, which naturally necessitates highly interpretable contextual examples~\citep{gao2023makes}. Similarly, code summarization~\citep{sukkasem2025llm,crupi2025effectiveness} aims to generate high-level natural language descriptions for source code. Existing studies emphasize that providing LLMs with specific semantic contexts or employing few-shot learning can significantly align the generated text with human intent~\citep{geng2024large,ahmed2024automatic}. However, generating summaries that strictly adhere to specific coding conventions requires project-specific context, an area that has not been systematically benchmarked.

During the software maintenance phase, LLMs are increasingly deployed for code review~\citep{zeng2025benchmarking,rasheed2024ai} and patch correctness assessment~\citep{zhang2026compass,yang2025parameter,zhou2024leveraging}. Code review is a critical yet labor-intensive process. While early learning-based approaches automated review comment generation~\citep{li2022auger}, recent empirical studies on LLMs (e.g., ChatGPT) reveal that effective code review requires complex reasoning and is heavily influenced by prompt designs and conversational contexts~\citep{guo2024exploring}. This highlights the necessity of modeling code review as a procedural decision-making process rather than a static classification task. Furthermore, Automated Program Repair (APR) often suffers from the patch overfitting issue. To assess patch correctness, researchers have explored utilizing LLMs for syntactic and semantic reasoning, transitioning from feature extractors to zero-shot or few-shot evaluators~\citep{zhou2023patchzero,le2023invalidator}. Distinguishing between valid and overfitting patches inherently demands comparative reasoning, making the exploration of positive and negative contexts highly relevant.

Despite the proliferation of LLM applications across these SE tasks, existing literature predominantly treats context selection as an ad-hoc component of prompt engineering. There remains a critical gap in systematically categorizing SE-specific context types and benchmarking their heterogeneous impacts on different SE tasks. This motivates the design of CL4SE, bridging the gap between general in-context learning and domain-specific SE practices.

\section{Benchmark}
Before presenting the details of \toolname{}, we compare it with representative existing benchmarks in Table~\ref{tab:comparison_with_benchmark}. Most prior SE benchmarks are task-specific, focusing on only one scenario. Moreover, they are typically limited to a single granularity level. Although several benchmarks have considered the role of context and demonstrated that contextual information can improve model performance, they do not provide a fine-grained analysis of how different categories of context influence different software engineering tasks. In contrast, \toolname{} unifies four representative SE tasks, spans multiple granularity levels, and supports systematic analysis of the relationship between context types and task-specific learning behavior.

\begin{table}[htbp]
  \centering
  \caption{Comparison between CL4SE and current benchmarks.}
    \resizebox{1.0\linewidth}{!}{\begin{tabular}{lccccc}
    \toprule
    Benchmark & \multicolumn{1}{l}{Time} & Language & Task  & Granularity & Context \\
    \midrule
    ClassEval~\citep{du2023classeval} & 2023  & Python & Code Generation & Class & No \\
    HumanEval & 2023  & Python & Code Generation & Function & No \\
    MBPP  & 2021  & Python & Code Generation & Function & No \\
    Code Contests & 2022  & Multilingual & Code Generation & File  & No \\
    SWEBench~\citep{DBLP:conf/iclr/JimenezYWYPPN24} & 2024  & Python & Code Generation & Commit & Yes \\
    ProjectEval~\citep{liu2025projecteval} & 2025  & Python & Code Generation & Project & Yes \\
    P-CodeSum~\citep{yun2024project} & 2024  & Multilingual & Code Summarization & Project & Yes \\
    ProConSuL~\citep{lomshakov2024proconsul} & 2024  & C/C++ & Code Summarization & Project & Yes \\
    ContextCRBench~\citep{hu2025benchmarking} & 2025  & Multilingual & Code Review & Commit & Yes \\
    AUGER~\citep{li2022auger} & 2022  & Java  & Code Review & Commit & Yes \\
    Defects4J~\citep{just2014defects4j} & 2014  & Java  & Code Repair \& PCA & Project & Yes \\
    Bugs.jar~\citep{saha2018bugs} & 2018  & Java  & Code Repair \& PCA & Project & Yes \\
    Bears~\citep{Madeiral2019} & 2019  & Java  & Code Repair \& PCA & Project & Yes \\
    CL-bench~\cite{dou2026cl} & 2026  & NL  & Multiple & - & Yes \\
    \midrule
    CL4SE (Ours) & 2026  & Multilingual & Multiple & Multiple & Yes \\
    \bottomrule
    \end{tabular}%
  \label{tab:comparison_with_benchmark}}%
\end{table}%

\subsection{Overview}
\toolname{} is designed to evaluate the learning ability and learning style of LLMs across different software engineering tasks and contexts. Unlike typical few-shot learning, which simply guides the model to better adapt to tasks by including examples in the prompt, \toolname{} focuses more on the types and organizational structures of contexts and their impact on the effectiveness of in-context learning for LLMs. Additionally, \toolname{} examines scenarios where the model may struggle to gain guidance from parameterized knowledge, particularly in niche or specialized contexts. This approach allows for a deeper understanding of how various contextual factors influence a model's performance and adaptability, ultimately leading to improved strategies for leveraging LLMs in diverse and complex software engineering tasks.

\begin{table}[htbp]
  \centering
  \caption{Statistics of \toolname{}, including tasks, contexts, data and evaluation.}
   \resizebox{1.0\linewidth}{!}{ \begin{tabular}{p{13em}lll}
    \toprule
    Task & Context  & Data  & Evaluation \\
    \midrule
    Code Generation & Interpretable examples &636   & Pass Rate \\
    Code Summarizatioon & Project-specific &8225  & ROUGE/BLEU/METEOR/BERTScore \\
    Code Review &  Procedural decision-making & 1916  & ACC/PRE/REC/F1 \\
    Patch Correctness Assessment & Positive \& negative & 2274  & ACC/PRE/REC/F1 \\
    \bottomrule
    \end{tabular}}%
  \label{tab:dataset_overview}%
\end{table}%

In \toolname{}, we select a total of four different software engineering tasks, including Code Generation (CG), Code Summarization (CS), Code Review (CR), and Patch Correctness Assessment (PCA). These tasks span various stages of software engineering and are both typical and representative. For each task, we collect a substantial amount of up-to-date data from real software projects and carefully design and select the contexts. In addition, \toolname{} is not limited to specific programming languages or types of LLMs. We select several mainstream programming languages, including Python, JavaScript, SQL, among others, and conduct evaluations across various LLMs, demonstrating the broad applicability of \toolname{}.
As shown in Table~\ref{tab:dataset_overview}, we present a summary of the data utilized in \toolname{}, including the volume of data for each task category, the contexts used by different tasks and the evaluation metrics employed.

\subsection{Context Taxonomy} 
We conduct an extensive investigation of diverse scenarios in software engineering and identify four representative categories of contexts. In what follows, we present the classification and semantics of each context type in turn.

\noindent\textbf{Project-specific context.} Software engineering often operates at the project level, and the software lifecycle consistently revolves around specific projects. In most SE tasks, contexts are generally shared across projects, such as code search~\citep{chen2024code,oskooei2025repository}, where semantic similarity is used to match code snippets. However, in certain tasks (e.g., code summarization~\citep{liu2024towards,yun2024project}, repository-level code completion~\citep{zhang2023repocoder,liu2024graphcoder}), the context exhibits project-specific characteristics.

The contexts of different projects can vary significantly; for instance, the style of code summarization may differ based on each project's objectives, coding standards, and team practices. This project-specificity greatly influences how effectively a model can generate relevant and coherent summaries that align with the unique coding conventions and terminologies of a particular project. However, modern LLMs do not differentiate between project affiliations in the code corpora during the pre-training phase. Instead, they typically mix all the code together for training. Consequently, the programming output from these models tends to lack project-specific characteristics, making it challenging to identify and leverage unique distinctions between projects. Therefore, it is essential to tailor contextual inputs for each project in order to enhance model performance in these specialized tasks. In this paper, we specifically distinguish this type of context to emphasize its importance and to examine the learning capabilities of LLMs in relation to it.

\noindent\textbf{Context based on interpretable examples.} Current few-shot learning methods mostly provide LLMs with superficial examples, such as output format examples and similar cases. For straightforward instruction-following tasks, such as requiring the model to output in JSON format, such examples can be quite effective. However, more complex tasks often require deeper explanations. LLMs may not be able to directly derive guidance from simple examples to apply to the current task. On one hand, for traditional non-reasoning models, it is very difficult to derive intermediate processes of a problem directly from the input-output pairs; if the examples are too dissimilar to the target problem, the model will lose meaningful guidance. On the other hand, for reasoning models, even if the model attempts to reason through the intermediate steps, it may fail or deviate from the expected path due to insufficient reasoning capabilities, thus introducing noise that interferes with solving the current task. In this paper, we aim to investigate the impact of the interpretable aspects of examples within the context on the model's ability to solve new problems. Therefore, we specifically distinguish this type of context to evaluate its effectiveness in enhancing the model's performance. By examining how interpretable examples influence problem-solving capabilities, we aim to provide insights that can guide the development of more effective learning strategies for complex tasks.

\noindent\textbf{Context based on positive \& negative examples.} Existing work typically uses positive examples to guide LLMs in learning the correct way to handle tasks~\citep{luo2023dr,rubin2022learning}, or negative examples to help the model avoid incorrect trajectories~\citep{liang2025failures,hamdan2025much}. However, there is limited research investigating the combined effects of these two different types of contexts, and even fewer studies compare the applicability and effectiveness of positive and negative examples across different tasks and models. Intuitively, positive and negative examples act like two distinct reference frames, and providing only one type may lead the model to focus excessively on that aspect. For instance, supplying only positive samples might lead the model to merely replicate the limited patterns within those examples, overlooking a broader range of possible solutions. Conversely, offering solely negative samples can teach the model how to avoid mistakes but does not guide it toward learning the correct approaches. This highlights the importance of a balanced approach that incorporates both types of examples, enabling the model to learn effectively by understanding both what to do and what to avoid. In this paper, we specifically distinguish between positive and negative contexts to compare their roles in specific software engineering tasks and the effects of their combination.

\noindent\textbf{Context based on procedural decision-making.} In addition to the three types of contexts mentioned above, we also distinguish a fourth type. We find that in some software tasks (e.g., code review), the determination of decisions (approve/disapprove) is not completed through a single interaction. Often, this interaction is a continuous and multi-turn process, during which the inclination of the decision can change. Previous relevant works~\citep{li2022auger,tufano2024code,frommgen2024resolving} have simplified the modeling by omitting the interactions and retaining only the final decision, neglecting the explainability and transparency of decision-making in software engineering tasks. This incomplete context can also lead to misjudgments by LLMs. In this paper, we focus on studying how to utilize procedural decision-making contexts to help LLMs understand and anticipate the perspectives and turning points that may arise throughout the decision-making process, ultimately leading to a well-reasoned decision.

\subsection{Task Selection}
\begin{figure*}[htbp]
  \centering
  \includegraphics[width=0.8\linewidth]{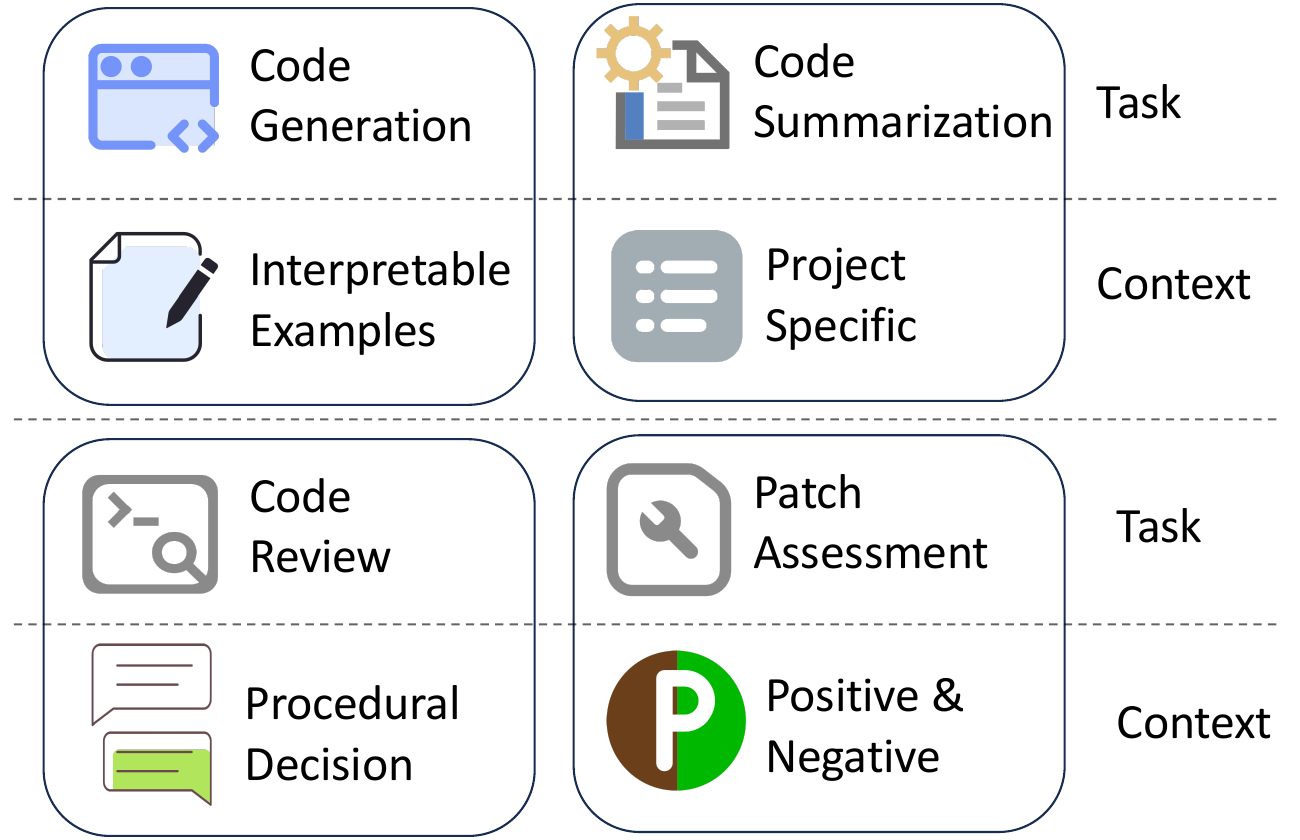}
  \caption{The mapping between context and task.} 
  \label{fig:task_context}
\end{figure*}

Against the four types of contexts proposed in the previous section, we further select four representative software engineering scenarios to examine the role and effectiveness of each context type in a task-specific setting. As shown in Figure~\ref{fig:task_context}, we present the mapping relationship between context types and downstream tasks, so as to provide a clearer view of how different forms of contextual information support different kinds of reasoning and decision-making.

For project-specific context, we choose code summarization, because this task strongly depends on project-level domain knowledge, implementation patterns, naming conventions, and coding styles that are often unique to a particular codebase. For context based on interpretable examples, we select code generation, since high-quality generation usually benefits from explicit reasoning demonstrations and transparent step-by-step guidance that can help the model better understand how to construct correct solutions. For context based on positive and negative examples, we adopt patch correctness assessment, as the core challenge of this task is to distinguish genuinely correct patches from overfitting or misleading ones, which naturally requires learning from both successful and unsuccessful cases. Finally, for context based on procedural decision, we apply it to code review, because code review often involves iterative analysis, multi-turn interaction, and progressive judgment before reaching a final conclusion. Through this one-to-one mapping between context types and tasks, we are able to systematically investigate how each type of context contributes to model performance in its most relevant software engineering scenario.

\subsection{Data Collection}
\subsubsection{Code Generation} We collected various types of problems from the well-known programming contest platform LeetCode. These problems cover a wide range of categories, including algorithm problems (e.g., two pointers, quicksort), data structure problems (e.g., linked lists, binary trees), real-world problem modeling, optimization, and so on. First, we crawled the problem descriptions and the official solution sets. Next, we collected various user discussions about solution approaches from the comment sections. Because these discussions contained noise, we further used Qwen3-Max to extract valuable and correct solution ideas from them. Because each problem description contains only up to three simple example cases, these examples alone are insufficient to judge the correctness of solutions. Therefore, we used the code model Qwen3‑Coder‑Plus to generate test cases. The requirements for model-generated cases were: (1) inputs and outputs should be diverse and non‑redundant; (2) include boundary cases whenever possible; (3) generate more than 10 cases per problem; and (4) the canonical solution must pass the generated cases. Through iterative generation, we first obtained a sufficient number of test cases, and then we manually evaluated and corrected the quality of the test cases. After filtering all valid samples (2,083 problems), we also rated problem difficulty. Using Qwen3‑32B as the baseline model, we ran repeated tests on the full dataset 10 times and used the average pass rate across the 10 runs as the metric to rank problems from lowest to highest. Finally, we selected the top 30\% as the test set (636 problems), the rest as the train set (1,447 problems).

\begin{table}[htbp]
  \centering
  \caption{Difficulty distribution of code generation benchmark.}
    \resizebox{0.6\linewidth}{!}{\begin{tabular}{l|cccc}
    \toprule
          & Hard  & Medium & Easy  & Total \\
    \midrule
    Train & 111   & 777   & 559   & 1447 \\
    Test  & 101   & 401   & 134   & 636 \\
    All   & 212   & 1178  & 693   & 2083 \\
    \bottomrule
    \end{tabular}}%
  \label{tab:difficulty}%
\end{table}%

The difficulty distributions for the train and test sets are shown in Table~\ref{tab:code_summary_result}. One surprising finding is that the model’s notion of difficulty does not always match the human labels in the LeetCode database: some problems labeled "hard" are easily solved by the model, while some "easy" problems are not. We believe this discrepancy is because classic problems that frequently appear in exams and technical interviews also occur often in various real‑world corpora; consequently, these problems were likely included in the model’s pretraining data and are therefore better mastered by the model. Therefore, by excluding problems with high pass rates from the test set, we can evaluate the model’s performance more objectively.

\begin{figure}[htbp]
\centering
    \subfigure[Tags Distribution in Train Set]{
        \includegraphics[width=0.45\columnwidth]{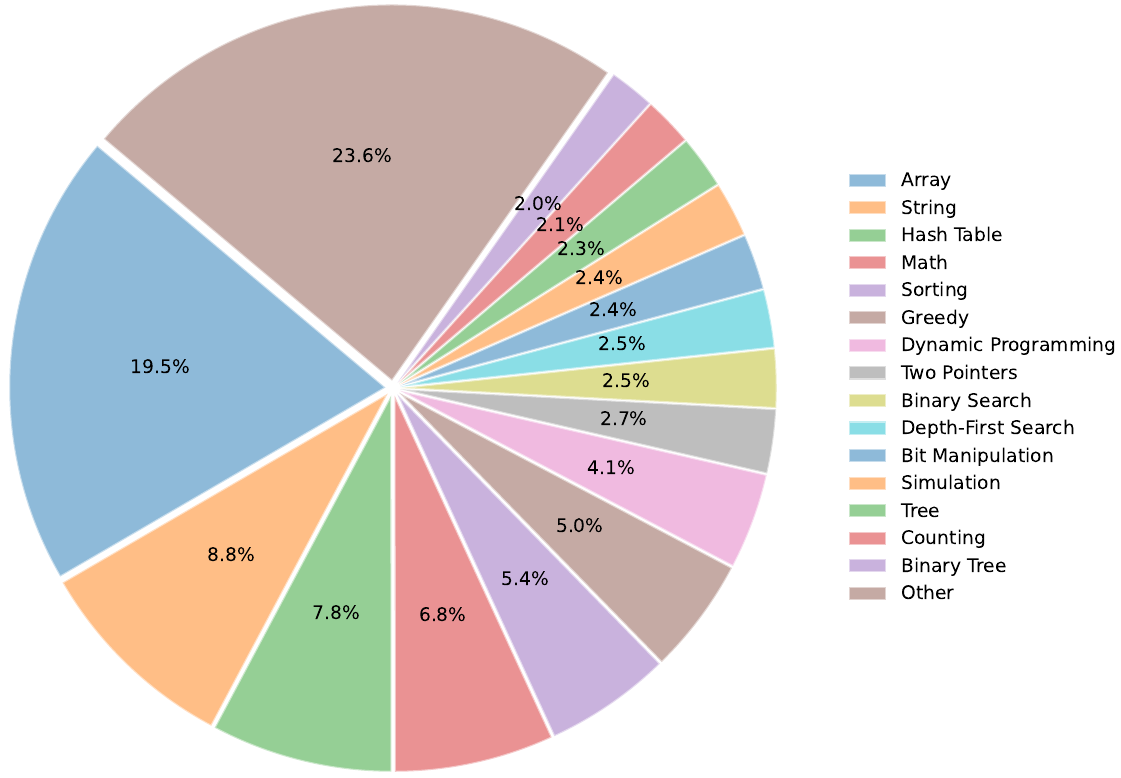}
        \label{fig:tag_train}
    }
    \subfigure[Tags Distribution in Test Set] {
        \includegraphics[width=0.45\columnwidth]{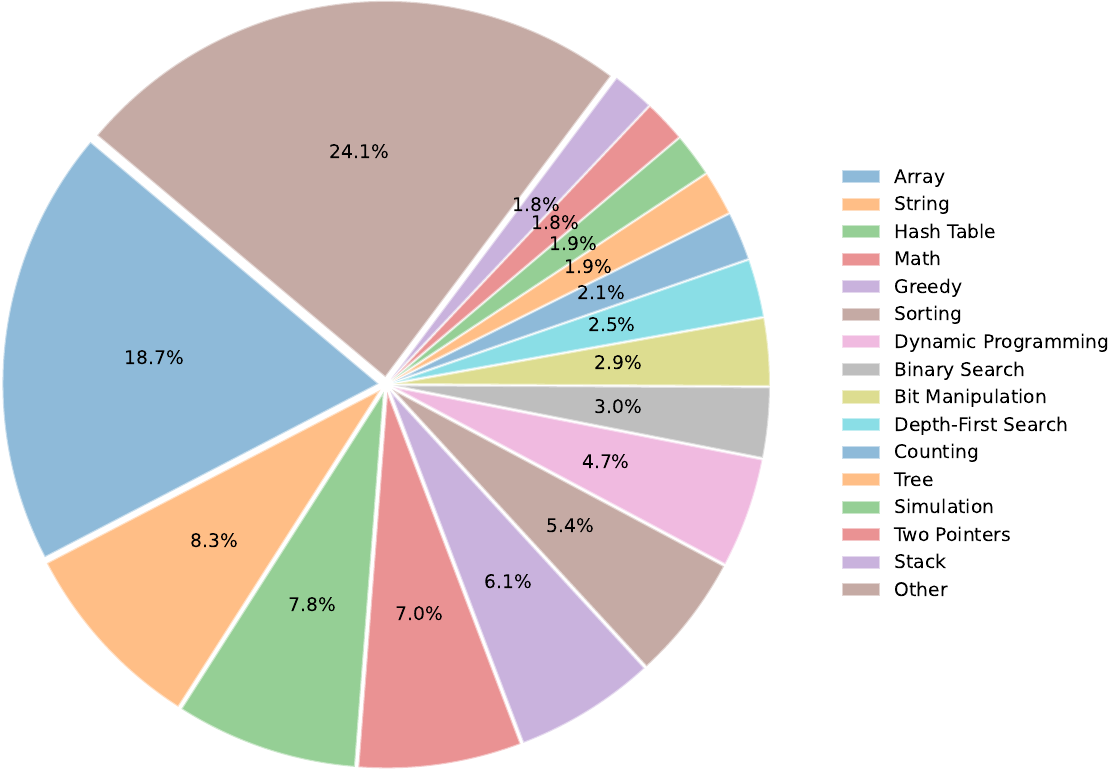}
        \label{fig:tag_test}
    }
\end{figure}

As shown in Figure~\ref{fig:tag_train} and Figure~\ref{fig:tag_test}, we also visualize the problem categories in the form of pie chart. The filtered test set does not concentrate on a few types of problems despite the reduced sample size; like the training set, it contains more than 15 categories. This demonstrates the richness and diversity of problem types in both sets of the code generation benchmark.

\subsubsection{Code Summarization}
To collect realistic code-summary samples from real-world software development, we aggregated the top ten large Python open-source projects on GitHub by star count. We cloned these repositories to our computing server and parsed each project at the function granularity using the general-purpose static analysis tool TreeSitter. For each extracted function we recorded metadata including project, file path, enclosing class (if any), function name, function body, and associated comments or docstrings. To improve the quality and relevance of the samples for summary tasks, we applied a set of conservative filtering rules: (1) exclude functions whose name or relative file path contains the substring "test" or "Test" to avoid test-only code; (2) exclude Python dunder methods (names that both start and end with "\_\_") since they typically represent language- or framework-level hooks rather than domain logic; (3) exclude functions that have no comments or docstrings, because our task requires a natural-language summary associated with executable code; and (4) exclude functions whose comment/docstring length exceeds 500 words to remove noisy or multi-paragraph documentation that does not reflect a concise summary. Any function that met one or more of these criteria was removed from the candidate pool.

After applying these filters and performing manual spot checks to verify filter correctness, we retained 8,225 high-quality samples (see Table~\ref{tab:code_summary_distribution}). These samples originate from 2,495 distinct source files and cover 1,197 unique classes, yielding approximately 3.3 samples per file and about 6.9 samples per class on average. The retained set includes a mixture of short utility functions and longer behavior-rich methods. The wide spread of files and classes indicates that the dataset is not dominated by a few modules or types, but rather provides broad coverage across repositories and object hierarchies. This diversity and the conservative filtering strategy together ensure that the dataset is both representative of real development code and well-suited for training and evaluating code-summary models.
\begin{table}[htbp]
  \centering
  \caption{Repo distribution of the code summarization benchmark.}
    \resizebox{0.7\linewidth}{!}{\begin{tabular}{lccc}
    \toprule
    Repo  & \# Files & \# Class & \# Summary \\
    \midrule
    AutoGPT & 201   & 102   & 824 \\
    LlamaFactory & 17    & 7     & 47 \\
    OpenHands & 165   & 70    & 333 \\
    django & 257   & 175   & 1060 \\
    fastapi & 14    & 5     & 35 \\
    flask & 18    & 13    & 73 \\
    langchain & 120   & 56    & 283 \\
    pytorch & 621   & 325   & 2445 \\
    transformers & 649   & 235   & 1903 \\
    vllm  & 433   & 209   & 1222 \\
    \midrule
    Total & 2495  & 1197  & 8225 \\
    \bottomrule
    \end{tabular}}%
  \label{tab:code_summary_distribution}%
\end{table}%

\subsubsection{Code Review}
To obtain realistic, high-quality code review data, we followed prior work such as SWE-bench-verified~\citep{DBLP:conf/iclr/JimenezYWYPPN24}, SWE-bench-lite~\citep{DBLP:conf/iclr/JimenezYWYPPN24}, and Multi-lingual SWE~\citep{DBLP:journals/corr/abs-2504-02605}, and preselected 32 well-known, high-star GitHub repositories. As shown in Table~\ref{tab:candidate_repo}, these repositories span multiple languages—including Python, JavaScript, and SQL—and cover diverse functionalities, ranging from component libraries to machine learning frameworks.
\begin{table}[htbp]
  \centering
  \caption{Candidate repos of the code review benchmark.}
    \resizebox{1.0\linewidth}{!}{\begin{tabular}{p{45.75em}}
    \toprule
    Candidate Project \\
    \midrule
    astropy/astropy, django/django, matplotlib/matplotlib, mwaskom/seaborn, pallets/flask, psf/requests, pydata/xarray, pylint-dev/pylint, pytest-dev/pytest, scikit-learn/scikit-learn, sphinx-doc/sphinx, sympy/sympy, sqlfluff/sqlfluff, marshmallow-code/marshmallow, pvlib/pvlib-python, pylint-dev/astroid, pyvista/pyvista, pydicom/pydicom, chartjs/Chart.js, grommet/grommet, openlayers/openlayers, eslint/eslint, scratchfoundation/scratch-gui, prettier/prettier, PrismJS/prism, highlightjs/highlight.js, GoogleChrome/lighthouse, bpmn-io/bpmn-js, alibaba-fusion/next, carbon-design-system/carbon, quarto-dev/quarto-cli, pytorch/pytorch, langchain/langchain \\
    \bottomrule
    \end{tabular}%
  \label{tab:candidate_repo}}%
\end{table}%

Second, we cloned the preselected GitHub repositories to our server. We then used the GitHub CLI to extract recent pull requests for each project. In terms of PR counts, we observed a large imbalance between Accepted and Rejected PRs: most projects have a majority of PRs approved by reviewers, while only a few projects (e.g., PyTorch) exhibit a comparatively high number of rejections. To ensure label balance in the dataset, we therefore sampled from each project’s recent PRs by selecting up to 5,00 accepted PRs and up to 5,000 rejected PRs.

We applied the following filtering criteria. First, we removed samples whose review conversation count is zero. Unlike prior code-review benchmarks that retain only a single review message, we found that reviews in real-world settings typically consist of multi-turn conversations involving multiple reviewers and the PR author. Because any single message cannot fully capture reviewers’ intentions, we preserve the entire review conversation. Second, we excluded PRs that modify more than three files to keep the dataset concise. Finally, we inspected the files changed by each PR and removed PRs that only modify test cases or that contain large, unrelated configuration-file changes ($\geq$ 10000 chars).

After applying the above filtering steps, as shown in Table~\ref{tab:code_review_dataset}, we obtained 1,916 valid code-review samples from the 32 candidate repositories, comprising 1,191 positive samples and 725 negative samples. To our knowledge, this is the benchmark with the largest number of projects and the closest alignment to real-world code review scenarios.

\begin{table}[htbp]
  \centering
  \caption{Repo distribution of the code review benchmark.}
    \resizebox{0.7\linewidth}{!}{\begin{tabular}{lccc}
    \toprule
    Project & \# Accept & \#Reject & \# Review \\
    \midrule
    lighthouse & 4     & 2     & 6 \\
    astropy & 148   & 44    & 192 \\
    bpmn-js & 1     & 1     & 2 \\
    carbon & 2     & 2     & 4 \\
    django & 239   & 9     & 248 \\
    eslint & 161   & 0     & 161 \\
    highlight.js & 3     & 1     & 4 \\
    matplotlib & 67    & 63    & 130 \\
    openlayers & 70    & 6     & 76 \\
    prettier & 23    & 8     & 31 \\
    requests & 51    & 4     & 55 \\
    xarray & 1     & 3     & 4 \\
    pydicom & 14    & 0     & 14 \\
    astroid & 41    & 12    & 53 \\
    pylint & 28    & 21    & 49 \\
    pytest & 43    & 33    & 76 \\
    pytorch & 9     & 476   & 485 \\
    pyvista & 19    & 14    & 33 \\
    quarto-cli & 47    & 1     & 48 \\
    scikit-learn & 112   & 19    & 131 \\
    sqlfluff & 108   & 6     & 114 \\
    \midrule
    Total & 1191  & 725   & 1916 \\
    \bottomrule
    \end{tabular}%
  \label{tab:code_review_dataset}}%
\end{table}%

\subsubsection{Patch Correctness Assessment}
Following prior work~\citep{tian2020evaluating,lin2022context}, we build our patch correctness assessment benchmark on Defects4J v2.0~\citep{just2014defects4j}, which contains 835 bugs across 17 open-source projects and remains the standard evaluation suite in APR and APCA research~\citep{zhang2023survey}. We aggregate plausible patches produced by recent APR tools and reported in the literature, including results from Liu et al.\cite{liu2020efficiency}, Xiong et al.\cite{xiong2018identifying}, Ghanbari et al.\cite{ghanbari2022patch}, Tian et al.\cite{tian2020evaluating,tian2022predicting,tian2022best}, and Lin et al.~\citep{lin2022context}, and we further mine released artifacts to locate additional plausible patches from current tools.

\begin{table}[htbp]
  \footnotesize
  \centering
  \caption{Statistics of patches in the patch correctness assessment benchmark.}
  \begin{tabular}{c|c|c|c|c}
    \toprule
    ID & APR Tool & \# Correct & \# Overfitting & \# Total \\
    \midrule
    1  & 3sFix        & 1    & 65   & 66   \\
    2  & ACS          & 32   & 13   & 45   \\
    3  & AVATAR       & 6    & 25   & 31   \\
    4  & Arja         & 16   & 187  & 203  \\
    5  & Arja-e       & 0    & 48   & 48   \\
    6  & CapGen       & 9    & 39   & 48   \\
    7  & Cardumen     & 0    & 8    & 8    \\
    8  & ConFix       & 4    & 58   & 62   \\
    9  & DeepRepair   & 4    & 8    & 12   \\
    10 & Developer    & 965  & 0    & 965  \\
    11 & DynaMoth     & 1    & 21   & 22   \\
    12 & Elixir       & 7    & 14   & 21   \\
    13 & FixMiner     & 3    & 26   & 29   \\
    14 & GenProg      & 1    & 26   & 27   \\
    15 & GenProgA     & 0    & 29   & 29   \\
    16 & HDRepair     & 7    & 2    & 9    \\
    17 & Hercules     & 0    & 4    & 4    \\
    18 & JGenProg2015 & 1    & 6    & 7    \\
    19 & Jaid         & 32   & 38   & 70   \\
    20 & Kali         & 0    & 15   & 15   \\
    21 & KaliA        & 3    & 40   & 43   \\
    22 & LSRepair     & 2    & 11   & 13   \\
    23 & Nopol        & 1    & 27   & 28   \\
    24 & Nopol2015    & 7    & 29   & 36   \\
    25 & Nopol2017    & 0    & 70   & 70   \\
    26 & PraPR        & 1    & 13   & 14   \\
    27 & RSRepair     & 0    & 8    & 8    \\
    28 & RSRepairA    & 4    & 34   & 38   \\
    29 & SOFix        & 2    & 1    & 3    \\
    30 & SequenceR    & 8    & 46   & 54   \\
    31 & SimFix       & 21   & 34   & 55   \\
    32 & SketchFix    & 3    & 8    & 11   \\
    33 & TBar         & 14   & 50   & 64   \\
    34 & genPat       & 0    & 1    & 1    \\
    35 & jGenProg     & 5    & 34   & 39   \\
    36 & jKali        & 3    & 20   & 23   \\
    37 & jMutRepair   & 0    & 12   & 12   \\
    38 & kPAR         & 3    & 30   & 33   \\
    39 & ssFix        & 3    & 5    & 8    \\
    \midrule
    \multicolumn{2}{c|}{Our Dataset} & 1169 & 1105 & 2274 \\
    \bottomrule
  \end{tabular}
  \label{tab:apca_datasets}
\end{table}

To mitigate label imbalance we also incorporate developer patches previously identified as correct~\citep{tian2020evaluating}. We then deduplicate the collected patches by reconstructing the original buggy and patched snippets, stripping whitespace and comments, and marking patches as duplicates when the normalized snippets are identical. Table~\ref{tab:apca_datasets} summarizes the resulting dataset statistics.

The outcome is a publicly shareable patch-assessment benchmark for Defects4J that, to our knowledge, is the largest of its kind: 2,274 plausible patches in total, of which 1,169 are overfitting patches and 1,105 are correct patches. This benchmark is intended to facilitate reproducible evaluation and comparison of patch-ranking and patch-classification techniques.

\subsection{Evaluation and Metrics}
In code generation tasks, we use PASS@1 to measure the correctness of the generated code. For code summarization tasks, we employ common text summarization and translation metrics (e.g., ROUGE~\citep{lin2004rouge}, BLEU~\citep{papineni2002bleu}, BERTScore~\citep{zhang2019bertscore}, METEOR~\citep{banerjee2005meteor}) to assess the quality of the generated summaries. In code review and patch correctness assessment tasks, we utilize Accuracy, Precision, Recall, and F1-score to evaluate the consistency between the model's generated answers and the ground truth. We provide a brief overview of the definitions of these metrics as follows.

\textit{PASS@1} evaluates whether the model's first output is successful in meeting the expected functionality, rather than requiring an exact match with the correct code.
Let \( N \) be the total number of code generation attempts and \( K \) be the number of attempts where the code passes the test cases on the first try. The PASS@1 success rate can be expressed as:
\begin{equation}
    \text{PASS@1} = \frac{K}{N}
\end{equation}
Where \( K \) is the count of successful first attempts that pass the test cases, \( N \) is the total number of code generation attempts.

\textit{BLEU}~\citep{papineni2002bleu} primarily measures the precision of n-grams (sequences of words) in the generated text compared with reference text.
\begin{equation}
    P_n = \frac{\text{Count of } n\text{-grams in generated text}}{\text{Count of } n\text{-grams in reference text}}
\end{equation}

\begin{equation}
    \text{BP} = 
\begin{cases} 
1 & \text{if } \text{length of generated} > \text{length of reference} \\
e^{(1 - \frac{\text{length of reference}}{\text{length of generated}})} & \text{otherwise}
\end{cases}
\end{equation}

\begin{equation}
    \text{BLEU} = BP \cdot \exp\left( \sum_{n=1}^{N} w_n \cdot \log(P_n) \right)
\end{equation}

Where \(w_n\) is a weight for each n-gram.

\textit{ROUGE}~\citep{lin2004rouge} measures the overlap between the generated summary and reference summaries.
\begin{equation}
    \text{ROUGE-P} = \frac{\text{Number of overlapping n-grams}}{\text{Total n-grams in generated summary}}
\end{equation}

\begin{equation}
    \text{ROUGE-R} = \frac{\text{Number of overlapping n-grams}}{\text{Total n-grams in reference summary}}
\end{equation}

\begin{equation}
    \text{ROUGE-F} = \frac{2 \cdot \text{ROUGE-P} \cdot \text{ROUGE-R}}{\text{ROUGE-P} + \text{ROUGE-R}}
\end{equation}

\textit{METEOR}~\citep{banerjee2005meteor} compares generated text against reference text by considering precision and recall, while also accounting for factors such as stemming, synonymy, and fragmentation penalties.

\begin{equation}
\text{F}_{\text{mean}} = \frac{2 \cdot P \cdot R}{P + R}
\end{equation}

\begin{equation}
    \text{METEOR} = \text{F}_{\text{mean}} \cdot (1 - \text{Penalty})
\end{equation}

Where \( P \) is the precision (correct matches divided by the total generated words), \( R \) is the recall (correct matches divided by the total reference words), Penalty accounts for fragmentation, reducing the score based on non-contiguous matches.

\textit{BERTScore}~\citep{zhang2019bertscore} evaluates the quality of generated text compared to reference text using embeddings derived from a pre-trained BERT model. By calculating precision, recall, and F1-score based on the similarity between the embeddings, it provides a comprehensive measure of textual similarity.

\begin{equation}
P_{BERT} = \frac{1}{|x|} \sum_{r_j \in x} \max_{\hat{y}_i \in \hat{x}} \text{sim}(r_j, \hat{y}_i)
\end{equation}

\begin{equation}
R_{BERT} = \frac{1}{|\hat{x}|} \sum_{\hat{y}_i \in \hat{x}} \max_{r_j \in x} \text{sim}(r_j, \hat{y}_i)
\end{equation}

\begin{equation}
F_{BERT} = 2 \cdot \frac{P_{BERT} \cdot R_{BERT}}{P_{BERT} + R_{BERT}}
\end{equation}

Where \( x \) is the embedding for the reference text, \( \hat{x} \) is the embeddings for the text generated by the model, and \( \text{sim}(r_j, \hat{y}_i) \) represents the similarity score, typically computed using cosine similarity.

\textit{Accuracy} measures the proportion of correctly reported samples.

\begin{equation}
Accuracy = \frac{TP + TN}{TP + TN + FP + FN}
\end{equation}

\textit{Recall} measures the ratio of reported positive samples over all the real positive samples.
\begin{equation}
Recall = \frac{TP }{TP + FN}
\end{equation}

\textit{Precision} measures the proportion of real positive samples over the reported positive samples.
\begin{equation}
Precision = \frac{TP }{TP + FP}
\end{equation}

\textit{F1} measures twice the multiplication of precision and recall divided by the sum of them.

\begin{equation}
F1 = 2 * \frac{Precision * Recall }{Precision + Recall}
\end{equation}

\section{Experiment}
\subsection{Research Questions}
In this paper, we study the effectiveness of context learning for software engineering tasks through the following research questions.

\noindent\textbf{RQ1:} What is the overall impact of context learning on different LLMs across various software engineering tasks?

\noindent\textbf{RQ2:} How does context learning influence the performance of LLMs on code generation with interpretable examples?

\noindent\textbf{RQ3:} To what extent does project-specific context learning affect code summarization quality?

\noindent\textbf{RQ4:} What is the effect of procedural decision context on code review judgment?

\noindent\textbf{RQ5:} How do positive and negative examples contribute to patch correctness assessment?

\subsection{Model Selection}
To evaluate context learning on software engineering tasks, we select 5 representative LLMs from diverse global manufacturers for comprehensiveness: Qwen3-Max~\citep{yang2025qwen3} and Qwen-Coder-Plus~\citep{bai2023qwen} (Alibaba Cloud, China), DeepSeek-V3~\citep{liu2024deepseek} (DeepSeek, China), GPT-Oss-120B~\citep{agarwal2025gpt} (OpenAI, US), and Claude-3.5-Haiku-20241022 (Anthropic, US). Covering both general and SE-specialized, closed-source and open-source models, these models ensure objective, generalizable evaluation, laying a foundation for our research questions.

\subsection{Implementation}
We present the implementation details of our experiments as follows, including model deployment, parameter configuration, search framework, and technical stack, ensuring reproducibility.
For model deployment, all experiments are conducted on an 8-card RTX 5880 GPU cluster. Closed-source models (Qwen3-Max, Qwen-Coder-Plus, DeepSeek-V3, Claude-3.5-Haiku-20241022) are accessed via their official APIs, with a concurrency request number set to 32 to balance efficiency and stability. The open-source model (GPT-Oss-120B) is deployed on 4 of the 8 RTX 5880 GPUs, using bfloat16 precision to optimize inference speed while maintaining output accuracy. All models are set with a temperature of 0 to ensure deterministic outputs without randomness.
The search framework is built using Langchain combined with the Chroma vector database, where the embedding model adopts Qwen3-Embedding-4B\footnote{https://huggingface.co/Qwen/Qwen3-Embedding-4B} to convert text and code samples into high-quality vector representations. During the search process, the target sample itself and samples with timestamps later than the target are excluded to avoid data leakage and ensure the fairness of experimental evaluations.
The entire experimental system is implemented based on the PyTorch and Transformer frameworks, ensuring efficient model inference, stable experimental operation, and convenient reproduction of experimental results by researchers.

\section{Main Results}
\subsection{RQ1: Overall Performances}
\noindent\textbf{Experiment design.} In RQ1, we focus on comparing the performance of LLMs under zero-shot and context learning settings across various software engineering tasks. For each task, we compute the average score across all evaluation metrics as the overall performance of a model on that task, and normalize it to the range [0, 1].

\begin{figure}[htbp]
\centering
    \subfigure[Code Generation]{
        \includegraphics[width=0.45\columnwidth]{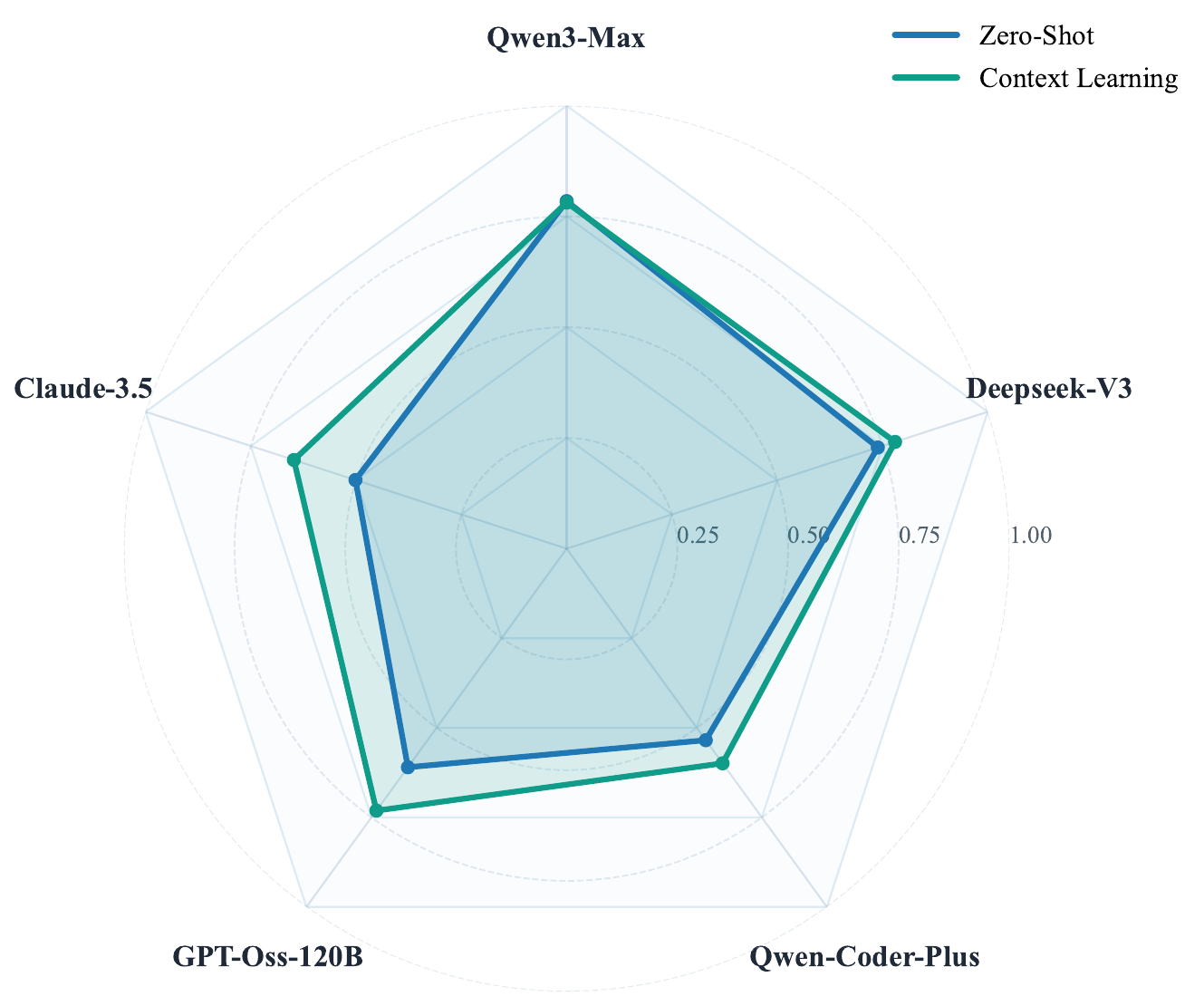}
        \label{fig:rayda_code_generation}
    }
    \subfigure[Code Summarization]{
        \includegraphics[width=0.45\columnwidth]{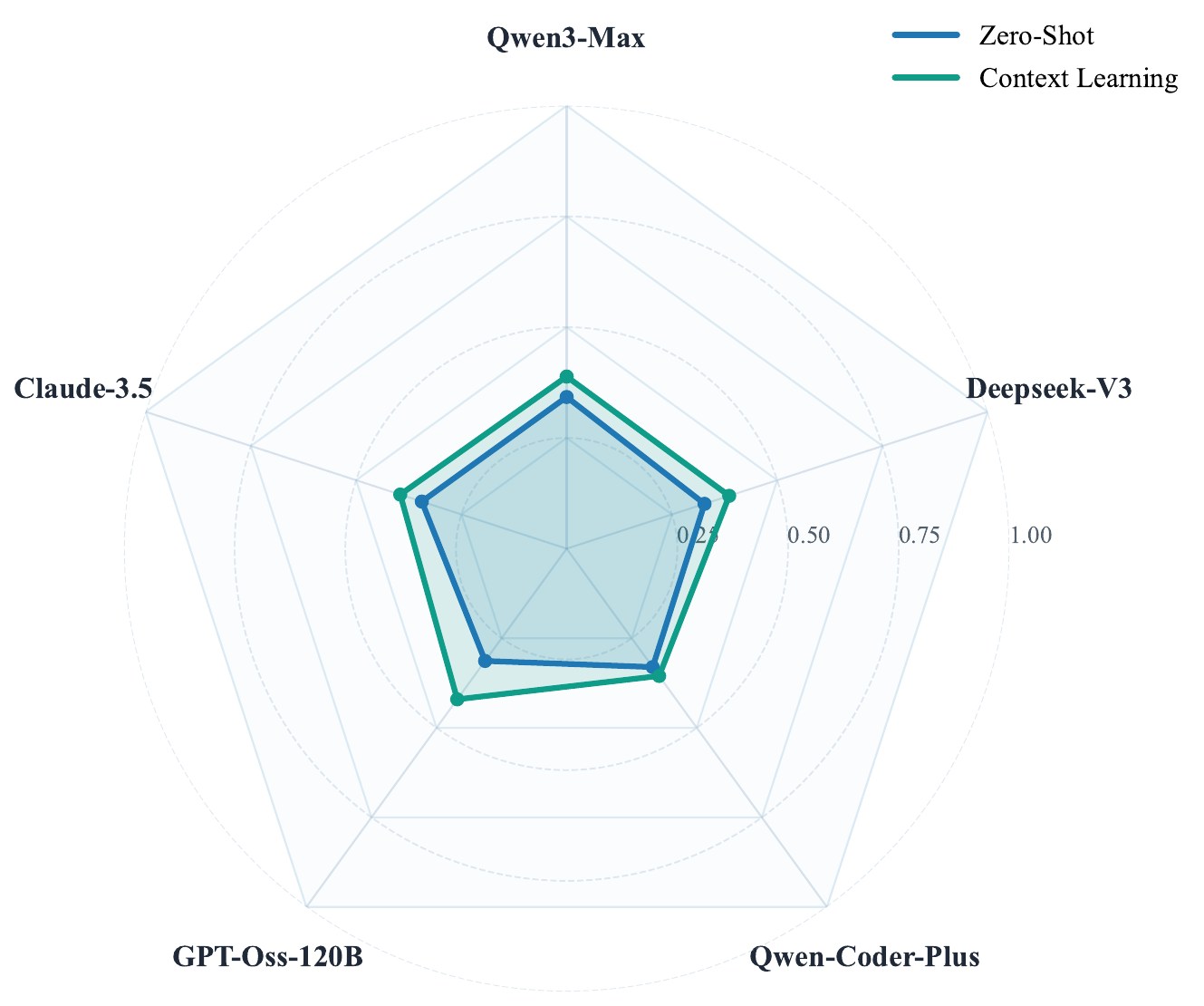}
        \label{fig:rayda_code_summarization}
    }
    \vspace{5pt} 

    \subfigure[Code Review] {
        \includegraphics[width=0.45\columnwidth]{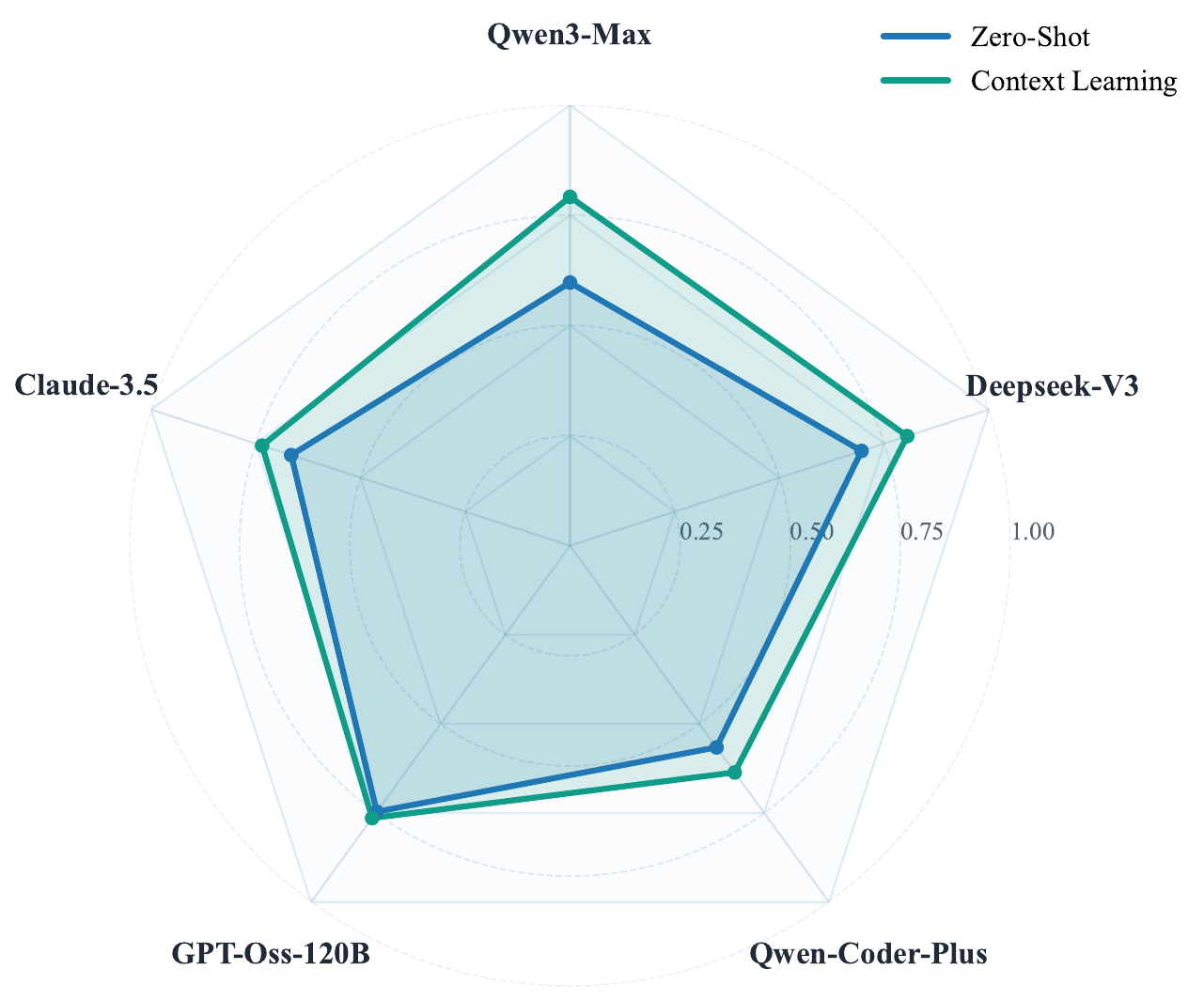}
        \label{fig:rayda_code_review}
    }
    \subfigure[Patch Correctness Assessment] {
        \includegraphics[width=0.45\columnwidth]{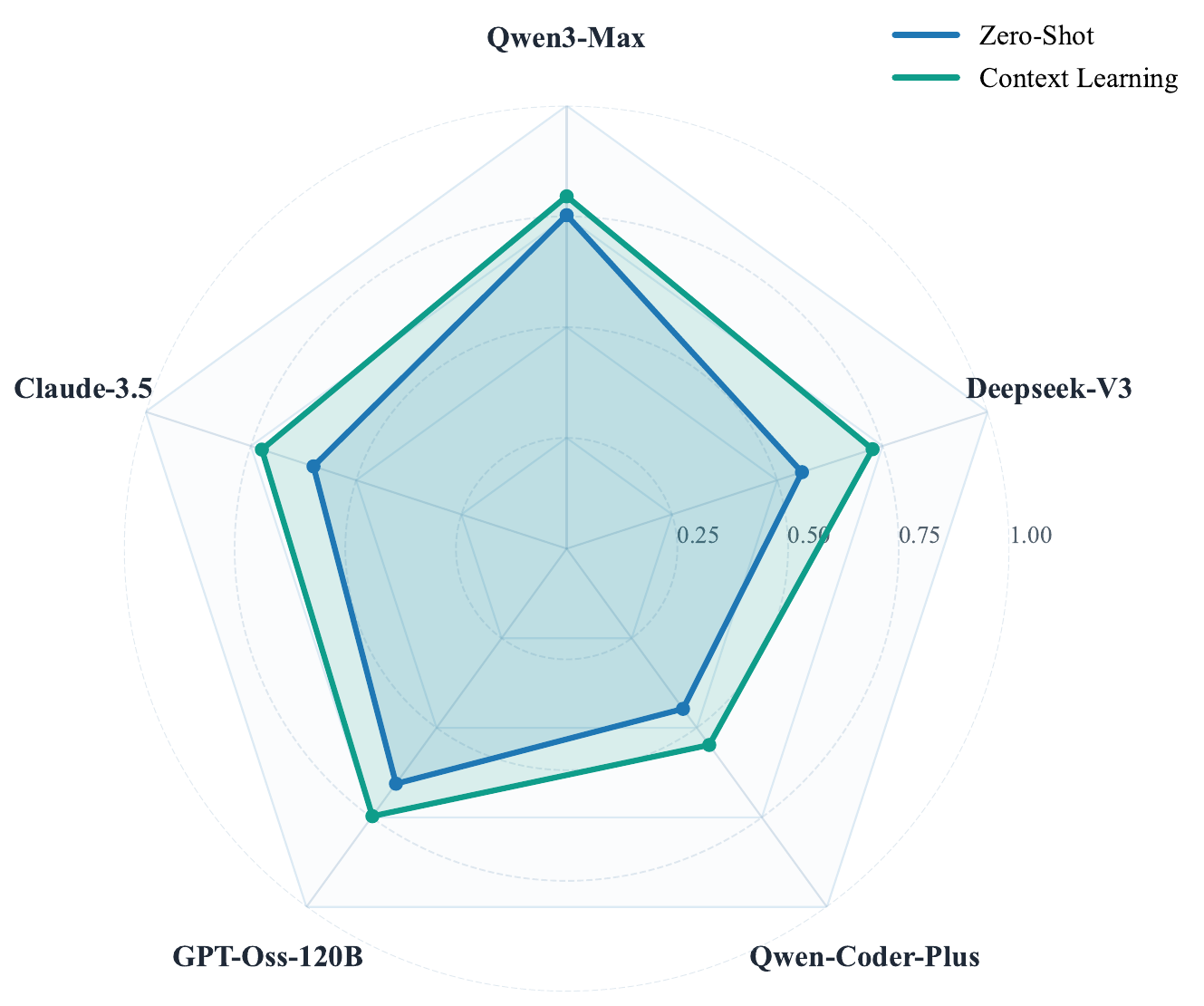}
        \label{fig:rayda_pca}
    }
\label{fig:rayda_all}
\end{figure}

\noindent\textbf{Overall Performance.} As shown in Figure~\ref{fig:rayda_code_generation} to Figure~\ref{fig:rayda_pca}, we use radar charts to compare the performance differences between context learning and zero-shot across the four tasks. Results show that context learning demonstrates a consistent superiority over zero-shot across nearly all models and tasks, confirming its general effectiveness in enhancing LLM performance for software engineering.

\noindent\textbf{Task and Model Specific Analysis.}
Across tasks, the magnitude of improvement varies significantly. Code Review sees the most substantial gains (e.g., Qwen3-Max improves by 33\%), followed by PCA (e.g., DeepSeek-V3 improves by 30\%). Code Generation and Code Summarization yield moderate but reliable improvements, with GPT-Oss-120B achieving the most notable lift in both tasks (20\% and 34\% respectively).
In terms of models, GPT-Oss-120B and Claude-3.5 exhibit the highest elasticity to context learning, leveraging contextual examples to narrow the gap with stronger baselines. Qwen3-Max, despite its already strong Zero-Shot performance, still benefits significantly from context learning in reasoning-intensive tasks (Code Review/PCA), with only a negligible drop in Code Generation due to a potential ceiling effect.

\finding{1}{

1. General Effectiveness: Context Learning is a robust paradigm that consistently improves LLM performance across diverse SE tasks, including CG, CS, CR and PCA.

2. Reasoning Sensitivity: Tasks demanding complex logical reasoning and decision-making (Code Review, PCA) benefit far more from context learning than tasks dominated by pattern matching.

3. Heterogeneous Adaptation: Models with lower Zero-Shot baselines (e.g., GPT-Oss-120B) achieve larger proportional gains from context learning compared to SOTA models like Qwen3-Max.
}

\subsection{RQ2: Context Learning on Code Generation}
\noindent\textbf{Experimental Design.} RQ2 aims to systematically investigate the impact of context learning with interpretable examples. We select code generation as the evaluation scenario and task. We first use examples without interpretations as context, allowing the model to directly learn the problem-solving logic of the current task from the problem description, target input-output pairs, and solutions. We then employ the same examples but add interpretations to illustrate the reasoning process and problem-solving strategies. By comparing the results under these two setups, we analyze the impact of interpretable examples on the model’s code generation capability.
In addition, we further explore how the number of shots affects model performance across different levels of problem difficulty.

\noindent\textbf{Results and Analysis.}
From the ablation results of interpretable examples in Table \ref{tab:ablation_analysis}, it can be clearly observed that the introduction of interpretability in contextual examples consistently yields positive improvements in the code generation PASS@1 rate for all models compared with the non-interpretation setting, which verifies the effectiveness of interpretable examples in enhancing the context learning ability of LLMs for code generation tasks. Specifically, there are significant heterogeneous differences in the magnitude of performance improvement across different models: DeepSeek-V3 achieves the most remarkable lift with an increase of 5.72\%, followed by Claude-3.5 (3.31\%) and GPT-Oss-120B (3\%), while Qwen-Coder-Plus and Qwen3-Max obtain relatively marginal gains of only 0.63\% and 0.15\% respectively. This result reveals a notable heterogeneity in model sensitivity to example interpretability: code-specialized models and general models with lower zero-shot baselines show significantly higher responsiveness to interpretable contexts, with their performance improved by 3\% to 5.72\% after adding interpretive reasoning. By contrast, Qwen3-Max, which already achieves a high zero-shot PASS@1 rate of 78.46\%, only gains a marginal 0.15\% improvement from interpretable examples, and even underperforms in the non-interpretation setting. This indicates that high-performance models have encoded sufficient code generation logic in their parameterized knowledge, making simple input-output examples unable to provide effective guidance and even likely to introduce noise, while interpretable contexts can effectively make up for the reasoning and task adaptation limitations of low-baseline models.

\begin{table}[htbp]
  \centering
   \caption{Effectiveness of interpretable examples on code generation task.}
    \resizebox{1.0\linewidth}{!}{\begin{tabular}{l|ccccc}
    \toprule
          & Qwen3-Max & DeepSeek-V3 & Qwen-Coder-Plus & GPT-Oss-120B & Claude-3.5 \\
    \midrule
    0-shot & 78.46\% & 73.90\% & 53.46\% & 61.01\% & 50.16\% \\
    w/o interpretation & 77.99\% & 72.27\% & 59.28\% & 70.12\% & 61.47\% \\
    w/ interpretation & 78.14\%(\textcolor{red}{$\uparrow$0.15\%}) & 77.99\%(\textcolor{red}{$\uparrow$5.72\%}) & 59.91\%(\textcolor{red}{$\uparrow$0.63\%}) & 73.12\%(\textcolor{red}{$\uparrow$3\%}) & 64.78\%(\textcolor{red}{$\uparrow$3.31\%}) \\
    \bottomrule
    \end{tabular}%
  \label{tab:ablation_analysis}}%
\end{table}%

Table \ref{tab:code_generation_result} further explores the impact of shot number (1–5 shot) on code generation performance across difficulty levels, using interpretable examples as the contextual setting. The results clearly demonstrate that performance does not scale linearly with the number of shots; instead, it exhibits a task-dependent optimal point and significant cross-difficulty heterogeneity.
First, across all models, 1-shot and 2-shot settings yield the peak performance on the overall dataset, with performance gradually declining as the shot number increases to 3–5 shot. DeepSeek-V3 achieves its highest overall PASS@1 rate at 1-shot (77.99\%), while GPT-Oss-120B is an exception, peaking at 4-shot (73.11\%) with a steady upward trend. In contrast, Qwen3-Max, Qwen-Coder-Plus, and Claude-3.5 all reach their maximum performance at 1-shot or 2-shot, followed by a noticeable drop in 3–5 shot settings, indicating that excessive demonstrations introduce redundant information and lead to performance degradation.
Second, the effectiveness of shot number is strongly tied to problem difficulty. 

On Easy problems, most models show a ceiling effect: Qwen3-Max maintains about 83\% performance across all shots, and Qwen-Coder-Plus only marginally improves from 58.96\% (0-shot) to 72.39\% (5-shot). On Medium problems, 1–2 shot settings consistently outperform higher shots: DeepSeek-V3 (77.81\% at 1-shot), GPT-Oss-120B (74.06\% at 4-shot), and Claude-3.5 (63.59\% at 1-shot) all achieve their best results in the low-shot range, confirming that interpretable examples are most impactful on practical, medium-difficulty tasks. On Hard problems, performance is unstable across shots with no clear optimal value: GPT-Oss-120B peaks at 5-shot (65.35\%), while Qwen3-Max and Claude-3.5 show erratic fluctuations, reflecting that even interpretable few-shot learning cannot fully mitigate the complexity of hard code generation problems.

\begin{table}[htbp]
  \centering
  \caption{Evaluating LLMs on code generation task.}
    \resizebox{0.8\linewidth}{!}{\begin{tabular}{lcccccc}
    \toprule
    \rowcolor[rgb]{ .906,  .902,  .902} Model & 0-shot & 1-shot & 2-shot & 3-shot & 4-shot & 5-shot \\
    \midrule
    Qwen3-Max & 78.46\% & 76.73\% & 78.14\% & 75.47\% & 75.00\% & 76.10\% \\
    Easy (134) & 83.58\% & 81.34\% & 83.58\% & 81.34\% & 83.58\% & 83.58\% \\
    Medium (401) & 78.80\% & 76.81\% & 78.30\% & 74.56\% & 73.32\% & 76.00\% \\
    Hard (101) & 70.30\% & 70.30\% & 70.30\% & 71.29\% & 70.30\% & 67.33\% \\
    \midrule
    DeepSeek-V3 & 73.90\% & 77.99\% & 77.04\% & 75.94\% & 75.79\% & 75.16\% \\
    Easy (134) & 75.37\% & 79.10\% & 78.36\% & 79.85\% & 74.63\% & 76.87\% \\
    Medium (401) & 72.32\% & 77.81\% & 77.06\% & 74.56\% & 75.56\% & 74.31\% \\
    Hard (101) & 78.22\% & 77.23\% & 75.25\% & 76.24\% & 78.22\% & 76.24\% \\
    \midrule
    Qwen-Coder-Plus & 53.46\% & 59.91\% & 59.91\% & 58.65\% & 58.96\% & 59.91\% \\
    Easy (134) & 58.96\% & 69.40\% & 68.66\% & 69.40\% & 67.16\% & 72.39\% \\
    Medium (401) & 51.12\% & 56.86\% & 58.10\% & 56.11\% & 56.11\% & 57.36\% \\
    Hard (101) & 55.45\% & 59.41\% & 55.45\% & 54.46\% & 59.41\% & 53.47\% \\
    \midrule
    GPT-Oss-120B & 61.01\% & 69.34\% & 71.70\% & 71.23\% & 73.11\% & 72.80\% \\
    Easy (134) & 57.46\% & 73.88\% & 76.87\% & 78.36\% & 79.10\% & 77.61\% \\
    Medium (401) & 64.09\% & 70.07\% & 72.32\% & 71.82\% & 74.06\% & 73.07\% \\
    Hard (101) & 53.47\% & 60.40\% & 62.38\% & 59.41\% & 61.39\% & 65.35\% \\
    \midrule
    Claude-3.5 & 50.16\% & 64.78\% & 64.78\% & 62.89\% & 61.79\% & 61.16\% \\
    Easy (134) & 58.96\% & 73.13\% & 73.88\% & 74.63\% & 74.63\% & 72.39\% \\
    Medium (401) & 48.88\% & 63.59\% & 62.59\% & 59.35\% & 58.60\% & 57.86\% \\
    Hard (101) & 43.56\% & 58.42\% & 61.39\% & 61.39\% & 57.43\% & 59.41\% \\
    \bottomrule
    \end{tabular}}%
  \label{tab:code_generation_result}%
\end{table}%

\finding{2}{

1. Interpretable examples outperform non-interpretation settings for all models, with code-specialized and low zero-shot baseline models gaining far more significantly.

2. Interpretable shot number has a non-linear effect, with 1–2 shot being optimal for most models and excess shots causing performance degradation.

3. Problem difficulty shapes interpretability efficacy, with medium-difficulty tasks as the primary beneficiaries and easy/hard tasks showing limited gains.

4. Interpretable contexts compensate for reasoning limitations of low-baseline models, while high-performance models see marginal gains due to pre-encoded knowledge.
}

\subsection{RQ3: Context Learning on Code Summarization}
\noindent\textbf{Experimental Design.} RQ3 explores how project-specific context impacts LLMs’ code summarization performance, which requires strict adherence to a project’s unique terminology, documentation style, and naming conventions.
Using a within-subjects design, we leverage 8,225 code-summary pairs from 10 Python open-source projects (e.g., PyTorch, Transformers). For each target code snippet, demonstration examples are strictly sourced from the same repository to ensure project-aligned domain knowledge and linguistic style.
We evaluate performance across 0-shot to 5-shot settings, using 15 metrics (ROUGE-1/2/L, BLEU, METEOR, BERTScore) to comprehensively assess lexical similarity, semantic adequacy, and contextual alignment of generated summaries.

\noindent\textbf{Results and Analysis.}
Table \ref{tab:code_summary_result} demonstrates a clear and consistent pattern in how LLMs utilize project-specific context: 1-shot is the definitive optimal setting across all models and metrics, followed by a monotonic performance decline as the number of project-specific shots increases.
All models achieve their peak scores at 1-shot, with striking differences in metric improvements: Lexical metrics (ROUGE-1/2/L, BLEU) see dramatic surges, while semantic alignment (BERTScore) shows only modest gains. For instance, GPT-Oss-120B’s BLEU score jumps from 5.57\% (0-shot) to 20.35\% (1-shot) (+14.78\%), and its Rouge-L F1 surges by 13.71\% (18.96\% $\rightarrow$ 32.67\%), indicating that project-specific context effectively guides the model to adopt the target project’s unique terminology, phrasing, and documentation format. In contrast, its BERTScore F1 only increases by 2.05\% (85.97\% $\rightarrow$ 88.02\%), which confirms that LLMs already generate semantically accurate summaries in the 0-shot setting, project-specific context primarily optimizes format and linguistic consistency rather than core meaning. This aligns with the key observation that summaries remain functionally correct without project-specific context, but lack alignment with the project’s internal documentation norms.
Notably, performance degrades sharply and consistently as shot number exceeds 1-shot: all models show steady declines in ROUGE, BLEU, METEOR, and BERTScore metrics from 2-shot to 5-shot. For example, GPT-Oss-120B’s Rouge-L F1 plummets to 20.75\% at 5-shot, and Qwen3-Max’s falls to 20.85\%, both dropping below their respective 0-shot baselines. This trend is mirrored in lexical metrics (e.g., Qwen3-Max’s BLEU drops from 9.75\% at 1-shot to 10.46\% at 5-shot, failing to retain initial gains) and BERTScore (e.g., Claude-3.5’s BERTScore F1 falls from 86.86\% to 85.95\%), indicating that excessive project-specific examples introduce harmful redundancy that erodes both linguistic alignment and semantic stability.
Model heterogeneity is also observed: general-purpose models (GPT-Oss-120B, Claude-3.5) show a more pronounced "peak-and-decline" pattern in lexical metrics, as they rely heavily on the 1-shot context to adapt to project-specific formatting and terminology. Code-specialized models like Qwen-Coder-Plus, while also peaking at 1-shot, exhibit more modest lexical gains (+4.19\% BLEU, 7.12\% $\rightarrow$ 11.31\%) and smaller BERTScore improvements (+0.18\%), suggesting they have partial built-in familiarity with common code documentation patterns but remain susceptible to noise from excessive context.

\begin{table}[htbp]
  \centering
  \caption{Evaluating LLMs on code summarization task.}
    \resizebox{1.0\linewidth}{!}{\begin{tabular}{lcccccccccccccc}
    \toprule
    \rowcolor[rgb]{ .906,  .902,  .902} Model & \multicolumn{3}{c}{Rouge-1} & \multicolumn{3}{c}{Rouge-2} & \multicolumn{3}{c}{Rouge-L} & Bleu  & Meteor & \multicolumn{3}{c}{BERTScore} \\
    \midrule
    \rowcolor[rgb]{ .906,  .902,  .902} Qwen3-Max & r     & p     & f     & r     & p     & f     & r     & p     & f     & s      &  s     & p     & r     & f1 \\
    0-shot & 31.67\% & 24.52\% & 25.32\% & 10.92\% & 7.73\% & 8.20\% & 28.58\% & 21.96\% & 22.75\% & 8.09\% & 29.07\% & 86.83\% & 86.92\% & 86.80\% \\
    1-shot & 33.52\% & 32.14\% & 30.74\% & 18.33\% & 16.98\% & 16.66\% & 31.83\% & 30.43\% & 29.17\% & 9.75\% & 31.73\% & 88.01\% & 87.10\% & 87.49\% \\
    2-shot & 29.60\% & 28.57\% & 27.17\% & 15.79\% & 14.21\% & 14.13\% & 28.25\% & 27.11\% & 25.88\% & 13.65\% & 28.11\% & 87.53\% & 86.60\% & 87.01\% \\
    3-shot & 26.61\% & 26.19\% & 24.71\% & 13.22\% & 12.41\% & 12.21\% & 25.31\% & 24.83\% & 23.50\% & 12.08\% & 25.08\% & 87.22\% & 86.21\% & 86.66\% \\
    4-shot & 24.61\% & 24.34\% & 22.92\% & 11.89\% & 11.21\% & 11.10\% & 23.42\% & 23.06\% & 21.79\% & 11.05\% & 23.28\% & 86.91\% & 85.92\% & 86.36\% \\
    5-shot & 23.39\% & 23.49\% & 21.96\% & 11.15\% & 10.73\% & 10.53\% & 22.21\% & 22.25\% & 20.85\% & 10.46\% & 22.15\% & 86.80\% & 85.80\% & 86.24\% \\
    \midrule
    \rowcolor[rgb]{ .906,  .902,  .902} DeepSeek-V3 & r     & p     & f     & r     & p     & f     & r     & p     & f     & s      &  s     & p     & r     & f1 \\
    0-shot & 30.00\% & 21.96\% & 23.18\% & 9.56\% & 6.26\% & 6.82\% & 26.95\% & 19.54\% & 20.70\% & 6.68\% & 27.45\% & 86.24\% & 86.64\% & 86.37\% \\
    1-shot & 32.07\% & 31.69\% & 29.95\% & 17.29\% & 16.31\% & 15.91\% & 30.45\% & 29.99\% & 28.41\% & 15.40\% & 30.42\% & 88.11\% & 86.93\% & 87.45\% \\
    2-shot & 28.31\% & 29.38\% & 27.26\% & 15.01\% & 14.86\% & 14.38\% & 27.00\% & 27.87\% & 25.97\% & 14.23\% & 27.24\% & 87.87\% & 86.45\% & 87.10\% \\
    3-shot & 25.83\% & 27.39\% & 25.14\% & 13.26\% & 13.56\% & 12.96\% & 24.63\% & 26.05\% & 23.98\% & 13.03\% & 24.80\% & 87.64\% & 86.15\% & 86.84\% \\
    4-shot & 23.82\% & 25.39\% & 23.24\% & 12.02\% & 12.13\% & 11.72\% & 22.72\% & 24.14\% & 22.16\% & 11.78\% & 23.03\% & 87.34\% & 85.85\% & 86.54\% \\
    5-shot & 22.69\% & 24.41\% & 22.20\% & 11.20\% & 11.51\% & 11.02\% & 21.61\% & 23.19\% & 21.15\% & 11.06\% & 21.88\% & 87.18\% & 85.67\% & 86.37\% \\
    \midrule
    \rowcolor[rgb]{ .906,  .902,  .902} Qwen-Coder-Plus & r     & p     & f     & r     & p     & f     & r     & p     & f     & s      &  s     & p     & r     & f1 \\
    0-shot & 25.30\% & 26.69\% & 23.82\% & 8.24\% & 8.20\% & 7.46\% & 23.20\% & 24.37\% & 21.78\% & 7.12\% & 25.88\% & 87.52\% & 86.27\% & 86.82\% \\
    1-shot & 27.98\% & 28.38\% & 26.33\% & 12.99\% & 12.27\% & 11.85\% & 26.43\% & 26.64\% & 24.81\% & 11.31\% & 27.34\% & 87.64\% & 86.50\% & 87.00\% \\
    2-shot & 27.07\% & 27.80\% & 25.71\% & 13.72\% & 12.89\% & 12.60\% & 25.81\% & 26.31\% & 24.45\% & 12.15\% & 26.17\% & 87.59\% & 86.26\% & 86.86\% \\
    3-shot & 24.71\% & 25.48\% & 23.45\% & 11.89\% & 11.20\% & 10.92\% & 23.58\% & 24.18\% & 22.34\% & 10.53\% & 23.63\% & 87.26\% & 85.88\% & 86.51\% \\
    4-shot & 22.80\% & 23.43\% & 21.56\% & 10.85\% & 9.86\% & 9.80\% & 21.77\% & 22.22\% & 20.53\% & 9.37\% & 21.81\% & 86.93\% & 85.58\% & 86.19\% \\
    5-shot & 21.63\% & 22.42\% & 20.51\% & 10.05\% & 9.16\% & 9.07\% & 20.65\% & 21.25\% & 19.53\% & 8.62\% & 20.79\% & 86.79\% & 85.44\% & 86.06\% \\
    \midrule
    \rowcolor[rgb]{ .906,  .902,  .902} GPT-Oss-120B & r     & p     & f     & r     & p     & f     & r     & p     & f     & s      &  s     & p     & r     & f1 \\
    0-shot & 28.24\% & 20.10\% & 21.52\% & 7.45\% & 4.86\% & 5.33\% & 25.01\% & 17.68\% & 18.96\% & 5.57\% & 25.99\% & 85.53\% & 86.55\% & 85.97\% \\
    1-shot & 35.72\% & 36.07\% & 34.06\% & 21.08\% & 21.21\% & 20.42\% & 34.21\% & 34.56\% & 32.67\% & 20.35\% & 34.14\% & 88.65\% & 87.52\% & 88.02\% \\
    2-shot & 29.50\% & 30.08\% & 28.10\% & 15.84\% & 15.82\% & 15.28\% & 28.14\% & 28.66\% & 26.84\% & 15.24\% & 28.30\% & 87.74\% & 86.63\% & 87.13\% \\
    3-shot & 26.45\% & 27.31\% & 25.27\% & 13.55\% & 13.73\% & 13.15\% & 25.27\% & 26.05\% & 24.16\% & 13.17\% & 25.44\% & 87.36\% & 86.22\% & 86.73\% \\
    4-shot & 24.21\% & 24.77\% & 22.98\% & 11.75\% & 11.71\% & 11.35\% & 23.05\% & 23.59\% & 21.92\% & 11.35\% & 23.31\% & 86.93\% & 85.85\% & 86.34\% \\
    5-shot & 22.76\% & 23.60\% & 21.72\% & 10.82\% & 10.93\% & 10.52\% & 21.71\% & 22.50\% & 20.75\% & 10.47\% & 21.77\% & 86.79\% & 85.65\% & 86.16\% \\
    \midrule
    \rowcolor[rgb]{ .906,  .902,  .902} Claude-3.5 & r     & p     & f     & r     & p     & f     & r     & p     & f     & s      &  s     & p     & r     & f1 \\
    0-shot & 43.77\% & 17.70\% & 22.76\% & 15.63\% & 5.31\% & 6.99\% & 40.50\% & 16.17\% & 20.86\% & 6.57\% & 29.24\% & 83.55\% & 87.26\% & 85.29\% \\
    1-shot & 40.23\% & 27.61\% & 29.81\% & 22.23\% & 15.10\% & 16.36\% & 38.17\% & 26.27\% & 28.35\% & 15.15\% & 33.27\% & 86.47\% & 87.42\% & 86.86\% \\
    2-shot & 33.21\% & 28.09\% & 28.00\% & 17.61\% & 14.71\% & 14.96\% & 31.57\% & 26.74\% & 26.68\% & 14.72\% & 29.68\% & 87.01\% & 86.92\% & 86.90\% \\
    3-shot & 28.83\% & 25.44\% & 24.93\% & 13.98\% & 12.38\% & 12.37\% & 27.45\% & 24.25\% & 23.78\% & 12.38\% & 25.99\% & 86.68\% & 86.38\% & 86.48\% \\
    4-shot & 26.57\% & 23.60\% & 23.01\% & 12.39\% & 10.95\% & 11.01\% & 25.23\% & 22.43\% & 21.90\% & 11.09\% & 24.05\% & 86.37\% & 86.05\% & 86.15\% \\
    5-shot & 24.76\% & 22.61\% & 21.72\% & 11.26\% & 10.25\% & 10.17\% & 23.53\% & 21.51\% & 20.68\% & 10.19\% & 22.60\% & 86.23\% & 85.79\% & 85.95\% \\
    \bottomrule
    \end{tabular}}%
  \label{tab:code_summary_result}%
\end{table}%

\finding{3}{

1. Project-specific context yields optimal code summarization performance at the 1-shot setting, with lexical metrics (ROUGE, BLEU) showing dramatic gains and BERTScore improving modestly.

2. The gap between lexical and semantic metric improvements confirms that LLMs generate semantically correct summaries in zero-shot setting, while project-specific context primarily optimizes alignment with a project’s internal documentation format and terminology.

3. Increasing project-specific shots leads to monotonic performance degradation due to contextual redundancy.

4. General-purpose models benefit more from project-specific context in linguistic alignment than code-specialized models, which show more resilience but smaller relative gains.

5. The 1-shot optimal point highlights that "less is more" when leveraging project-specific context, which is sufficient to align with internal norms without introducing redundant noise.

}

\subsection{RQ4: Context Learning on Code Review}
\noindent\textbf{Experimental Design.} RQ4 focuses on exploring the impact of procedural decision-making context on LLMs’ code review judgment, which is an inherently multi-turn task where decisions (approve/reject) emerge through progressive discussions rather than single-step evaluations.
We adopt a controlled within-subjects design, leveraging 1,916 high-quality code review samples from 32 GitHub repositories (covering Python, JavaScript, SQL). Each sample retains the full multi-turn conversation history (reviewer comments, author responses, decision turning points) as procedural context, rather than only the final decision. We evaluate model performance across 0-shot to 5-shot settings. In each shot, the prompt includes complete procedural decision examples from the same repository to ensure consistency in review norms and discussion styles.

\noindent\textbf{Results and Analysis.} Table \ref{tab:code_review_result} reveals that procedural decision-making context consistently enhances code review performance, with distinct trends in model behavior and metric improvements that stand in sharp contrast to previous tasks.
First, all models exhibit a steady, monotonic upward trend in core metrics (Accuracy, F1) as the shot number increases from 0 to 5, with no evidence of performance degradation from excessive context. Qwen3-Max demonstrates the most dramatic growth, with Accuracy rising by 18.69\% (58.19\% $\rightarrow$ 76.88\%) and F1-score surging by 22.80\% (56.39\% $\rightarrow$ 79.19\%). DeepSeek-V3 and Claude-3.5 also show substantial gains, confirming that procedural context provides cumulative, additive value for decision-making tasks, where learning from more examples refines the model’s understanding of nuanced review criteria.
Second, metric heterogeneity highlights how context refines the precision-recall balance. For Qwen3-Max and Claude-3.5, the primary gain comes from a massive increase in Recall (Qwen3-Max: 43.21\% $\rightarrow$ 70.78\%; Claude-3.5: 56.02\% $\rightarrow$ 66.05\%), with Precision remaining stable or slightly improved. This indicates that procedural examples teach models to identify more true positive issues (flawed PRs) that they would otherwise miss in a 0-shot setting. DeepSeek-V3 shows a similar focus on Recall (65.41\% $\rightarrow$ 90.12\%), while GPT-Oss-120B uses the context to marginally boost Precision (64.05\% $\rightarrow$ 65.93\%), reducing its initial rate of false positives.
Notably, Qwen-Coder-Plus, despite steady improvement (Accuracy +12.04\%), remains the lowest-performing model across all settings. This suggests that code specialization alone is insufficient for code review; the task demands understanding of social negotiation and procedural norms in collaborative development, which general-purpose models (Qwen3-Max, Claude-3.5) adapt to more effectively with additional context. Finally, 5-shot emerges as the optimal setting, as the complexity of procedural decision-making requires exposure to a diverse set of discussion patterns and edge cases to reach peak performance.

\begin{table}[htbp]
  \centering
  \caption{Evaluating LLMs on code review task.}
    \resizebox{1.0\linewidth}{!}{\begin{tabular}{rlcccccc}
    \toprule
    \rowcolor[rgb]{ .906,  .902,  .902} \multicolumn{1}{l}{Model} & Metric & 0-shot & 1-shot & 2-shot & 3-shot & 4-shot & 5-shot \\
    \midrule
    \multicolumn{1}{l}{\multirow{4}[2]{*}{Qwen3-Max}} & Accuracy & 58.19\% & 64.09\% & 71.09\% & 74.06\% & 75.52\% & 76.88\% \\
          & Precision & 81.14\% & 86.29\% & 88.98\% & 89.52\% & 89.32\% & 89.87\% \\
          & Recall & 43.21\% & 50.21\% & 61.04\% & 65.99\% & 68.85\% & 70.78\% \\
          & F1    & 56.39\% & 63.48\% & 72.41\% & 75.98\% & 77.76\% & 79.19\% \\
    \midrule
    \multicolumn{1}{l}{\multirow{4}[2]{*}{DeepSeek-V3}} & Accuracy & 65.97\% & 69.83\% & 71.56\% & 73.08\% & 73.92\% & 75.09\% \\
          & Precision & 76.45\% & 72.39\% & 72.89\% & 73.73\% & 74.19\% & 74.93\% \\
          & Recall & 65.41\% & 83.21\% & 86.38\% & 88.06\% & 88.98\% & 90.12\% \\
          & F1    & 70.50\% & 77.42\% & 79.06\% & 80.26\% & 80.92\% & 81.83\% \\
    \midrule
    \multicolumn{1}{l}{\multirow{4}[2]{*}{Qwen-Coder-Plus}} & Accuracy & 48.80\% & 54.02\% & 56.11\% & 58.35\% & 58.40\% & 60.84\% \\
          & Precision & 77.19\% & 82.28\% & 81.26\% & 82.48\% & 86.20\% & 83.89\% \\
          & Recall & 44.18\% & 46.51\% & 47.61\% & 47.81\% & 48.11\% & 48.68\% \\
          & F1    & 56.20\% & 59.42\% & 60.05\% & 60.53\% & 61.75\% & 61.61\% \\
    \midrule
    \multicolumn{1}{l}{\multirow{4}[2]{*}{GPT-Oss-120B}} & Accuracy & 63.62\% & 64.77\% & 66.34\% & 65.60\% & 66.07\% & 66.39\% \\
          & Precision & 64.05\% & 64.44\% & 65.56\% & 65.09\% & 65.43\% & 65.93\% \\
          & Recall & 94.54\% & 96.64\% & 96.55\% & 96.30\% & 96.30\% & 95.04\% \\
          & F1    & 76.36\% & 77.32\% & 78.10\% & 77.68\% & 77.92\% & 77.85\% \\
    \midrule
    \multicolumn{1}{l}{\multirow{4}[2]{*}{Claude-3.5}} & Accuracy & 59.53\% & 62.19\% & 65.22\% & 65.81\% & 64.29\% & 67.72\% \\
          & Precision & 83.57\% & 83.36\% & 84.13\% & 83.39\% & 82.34\% & 85.61\% \\
          & Recall & 56.02\% & 57.71\% & 63.69\% & 65.59\% & 63.71\% & 66.05\% \\
          & F1    & 67.08\% & 68.20\% & 72.50\% & 73.43\% & 71.84\% & 74.57\% \\
    \bottomrule
    \end{tabular}%
  \label{tab:code_review_result}}%
\end{table}%

\finding{4}{

1. Procedural decision-making context steadily improves code review performance up to 5-shot, with no redundancy-induced degradation.

2. The context primarily boosts Recall (flawed PR detection) while stabilizing Precision, aligning with real-world review needs.

3. General-purpose models benefit more than code-specialized ones, as context learning both relies on procedural reasoning and code understanding.

}

\subsection{RQ5: Context Learning on Patch Correctness Assessment}
\noindent\textbf{Experimental Design.} RQ5 investigates how positive \& negative context impacts LLMs’ ability to assess patch correctness, which is an essential task requiring differentiation between valid patches (positive examples) and overfitting patches (negative examples).
We adopt a controlled within-subjects design, leveraging 2,274 patch samples from Defects4J v2.0 (1,105 correct patches, 1,169 overfitting patches) to ensure balanced labels. The experiment tests three context settings: (1) 1-shot negative (only overfitting patches), (2) 1-shot positive (only correct patches), and (3) 2-shot mixed (one correct + one overfitting patch). We use the same 5 LLMs as previous RQs and evaluate performance via Accuracy, Precision, Recall, and F1-score, focusing on how positive/negative examples independently and jointly guide patch classification.

\noindent\textbf{Results and Analysis.}
\begin{table}[htbp]
  \centering
  \caption{Evaluating LLMs on patch correctness assessment task.}
    \resizebox{1.0\linewidth}{!}{\begin{tabular}{lcccc}
    \toprule
    \rowcolor[rgb]{ .906,  .902,  .902} Qwen3-Max & 0-shot & 1-shot(Overfit) & 1-shot(Correct) & 2-shot(Correct+Overfit) \\
    Accuracy & 76.72\% & 74.48\% & 79.36\% & 80.41\% \\
    Precision & 94.19\% & 94.41\% & 92.06\% & 92.18\% \\
    Recall & 58.25\% & 53.46\% & 65.44\% & 67.58\% \\
    F1    & 71.99\% & 68.27\% & 76.50\% & 77.99\% \\
    \midrule
    \rowcolor[rgb]{ .906,  .902,  .902} DeepSeek-V3 & Vanilla & 1-shot(Overfit) & 1-shot(Correct) & 2-shot(Correct+Overfit) \\
    Accuracy & 61.53\% & 69.30\% & 70.93\% & 74.48\% \\
    Precision & 92.71\% & 89.70\% & 90.95\% & 89.41\% \\
    Recall & 27.20\% & 45.42\% & 48.16\% & 57.06\% \\
    F1    & 42.06\% & 60.31\% & 62.98\% & 69.66\% \\
    \midrule
    \rowcolor[rgb]{ .906,  .902,  .902} Qwen-Coder-Plus & Vanilla & 1-shot(Overfit) & 1-shot(Correct) & 2-shot(Correct+Overfit) \\
    Accuracy & 54.02\% & 55.16\% & 59.95\% & 60.74\% \\
    Precision & 93.57\% & 92.05\% & 96.06\% & 93.10\% \\
    Recall & 11.21\% & 13.86\% & 22.93\% & 25.41\% \\
    F1    & 20.02\% & 24.09\% & 37.02\% & 39.92\% \\
    \midrule
    \rowcolor[rgb]{ .906,  .902,  .902} GPT-Oss-120B & Vanilla & 1-shot(Overfit) & 1-shot(Correct) & 2-shot(Correct+Overfit) \\
    Accuracy & 68.85\% & 72.93\% & 74.82\% & 76.19\% \\
    Precision & 87.20\% & 91.69\% & 89.30\% & 90.86\% \\
    Recall & 46.06\% & 51.97\% & 57.88\% & 59.59\% \\
    F1    & 60.28\% & 66.34\% & 70.23\% & 71.98\% \\
    \midrule
    \rowcolor[rgb]{ .906,  .902,  .902} Claude-3.5 & Vanilla & 1-shot(Overfit) & 1-shot(Correct) & 2-shot(Correct+Overfit) \\
    Accuracy & 64.51\% & 72.11\% & 71.94\% & 74.26\% \\
    Precision & 94.57\% & 94.80\% & 94.17\% & 94.23\% \\
    Recall & 32.76\% & 48.33\% & 48.33\% & 53.12\% \\
    F1    & 48.67\% & 64.02\% & 63.88\% & 67.94\% \\
    \bottomrule
    \end{tabular}}%
  \label{tab:pca_result}%
\end{table}%

Table \ref{tab:pca_result} reveals three core findings about positive \& negative context.
First, positive examples outperform negative examples alone across all models. For Qwen3-Max, 1-shot correct (Accuracy=79.36\%, F1=76.50\%) outperforms 1-shot overfit (74.48\%, 68.27\%); similarly, DeepSeek-V3’s 1-shot correct (70.93\%, 62.98\%) surpasses 1-shot overfit (69.30\%, 60.31\%). This indicates that learning "what a valid patch looks like" provides more direct guidance than only knowing "what to avoid."
Second, the 2-shot mixed setting (correct + overfit) yields the best overall performance for all models. DeepSeek-V3 achieves the largest gains (Accuracy=74.48\%, F1=69.66\%), a 12.95\% F1 lift over the vanilla setting; Qwen3-Max (80.41\%, 77.99\%) and GPT-Oss-120B (76.19\%, 71.98\%) also reach peak performance in this setup. This confirms that combining positive and negative examples helps LLMs establish dual reference frames, thus understanding both valid patterns and invalid pitfalls.
Third, model heterogeneity is pronounced: code-specialized models (Qwen-Coder-Plus) show the smallest absolute gains (F1 +19.90\% from vanilla to 2-shot) but retain high Precision ($\geq$ 93\%), while general-purpose models (DeepSeek-V3, Claude-3.5) achieve larger F1 improvements by balancing Precision and Recall. Notably, all models maintain high Precision ($\geq$ 89\%) across settings, but Recall improves drastically with context, indicating that examples primarily help models identify more true valid/overfitting patches without sacrificing correctness.

\finding{5}{

1. Positive examples alone outperform negative examples in guiding patch correctness assessment.

2. The 2-shot mixed setting (correct + overfit) achieves optimal performance, confirming complementary value of dual reference frames.

3. Context primarily boosts Recall while preserving high Precision, with general-purpose models gaining more than code-specialized ones.
}

\section{Discussion}

While our work demonstrates the effectiveness of context learning for enhancing LLMs’ performance on mainstream SE tasks, several critical challenges remain unaddressed and warrant further investigation for real-world adoption of SE context learning. Below we highlight two prominent challenges that pose significant barriers to practical deployment.

1). \textbf{Context Learning for Low-Resource Programming Languages.}
Most existing SE context learning research focuses on high-resource languages (e.g., Python) with abundant open-source code, documentation, and task-specific data. 
However, low-resource programming languages (e.g., Rust, Julia, or domain-specific languages like Solidity for smart contracts, Verilog for hardware design) present unique challenges for context learning.
Due to the scarcity of high-quality contextual examples, low-resource languages lack sufficient task-specific context, which limits the construction of effective in-context demonstrations. Their unique linguistic and syntactic characteristics require strict alignment with language-specific semantics, which generic context engineering strategies cannot satisfy. Moreover, LLMs pre-trained on high-resource languages suffer from inherent cross-language transfer gaps, making it difficult to generalize context learning capabilities across different programming languages.This challenge is especially critical for niche industrial domains that rely on low-resource languages, where context learning must adapt to specialized language constraints rather than generic SE patterns.

2). \textbf{Dynamic and Evolving Project Contexts.}
SE context learning assumes static project contexts, where coding conventions, documentation styles, and decision norms are fixed during data collection. However, real-world software projects are dynamic and evolving, posing three key challenges:

\begin{itemize}[leftmargin=*]
    \item Context drift: Project contexts change over time, making static examples outdated or misleading.
    \item Long-tail context scarcity: Emerging edge cases lack corresponding historical contextual examples.
    \item Context scalability: Large projects contain fragmented sub-contexts, greatly increasing the difficulty of context construction and maintenance.
\end{itemize}

Above reveals a key limitation of current context learning methods: they perform well on static projects but cannot adapt to the evolving nature of software development. Future research should explore dynamic context engineering mechanisms to maintain alignment between contextual inputs and real-time project states.

\section{Conclusion}
This work addresses the lack of systematic evaluation for LLM context learning in software engineering by proposing \toolname{}, the first dedicated context learning benchmark for the SE domain. We define a fine-grained taxonomy of four SE-specific context types and map each to a representative core SE task. We construct a high-quality real-world dataset with 13,000+ samples from over 30 open-source projects (636 code generation problems, 8,225 code summarization pairs, 1,916 code review PRs, 2,274 patch assessment samples) and conduct extensive evaluations on five mainstream LLMs using 9 metrics to dissect the efficacy of different context types.

Extensive experiments confirm that context learning is a robust paradigm for SE tasks, delivering an average 24.7\% performance improvement across all tasks with clear task-specific optimal strategies: 1–2 shot interpretable examples boost DeepSeek-V3’s code generation PASS@1 by 5.72\%; 1-shot project-specific context raises GPT-Oss-120B’s code summarization BLEU by 14.78\% (from 5.57\% to 20.35\%); 5-shot procedural context increases Qwen3-Max’s code review Accuracy and F1 by 18.69\% and 22.80\% respectively; and 2-shot mixed positive-negative context lifts DeepSeek-V3’s patch assessment F1 by 12.95\%. General-purpose models benefit more from context learning than code-specialized ones in most decision-centric and linguistic alignment tasks.

We also identify key open challenges for real-world SE context learning deployment, including the scarcity of high-quality context examples for low-resource programming languages and the inability of static context design to adapt to the dynamic evolution of software projects (context drift, long-tail context scarcity, context scalability), as discussed in our analysis. These challenges point to critical future research directions in dynamic context engineering and cross-language context transfer for low-resource SE domains.

In summary, \toolname{} makes three core contributions: first, it establishes a standardized evaluation framework for SE-focused context learning with a well-defined context taxonomy and task-context mapping; second, it releases a large-scale, rigorously constructed dataset to facilitate reproducible research and model comparison; third, it provides actionable empirical insights into task-specific context design, demonstrating that context engineering is not a one-size-fits-all solution and offering concrete optimal strategies for core SE workflows. Our findings enable researchers and practitioners to move beyond ad-hoc prompt crafting and leverage principled context design to fully unlock LLMs’ potential in real-world software development. 
\section*{Declarations}

\noindent \textbf{Ethical approval}: Not Applicable.

\noindent \textbf{Informed consent}: Not Applicable.

\noindent \textbf{Author Contributions}.
Haichuan Hu: Conceptualization, Methodology, Writing - original draft, Investigation.
Quanjun Zhang: Conceptualization, Writing review \& editing, Validation.
Ye Shang: Software, Data curation, Writing - review \& editing. 
Guoqing Xie: Software, Formal analysis, Writing - review \& editing.
Chunrong Fang: Conceptualization, Writing review \& editing, Validation.
Zhenyu Chen: Conceptualization, Writing review \& editing, Validation.
Liang Xiao: Conceptualization, Writing review \& editing, Validation.

\noindent \textbf{Data Availability Statement}.
Our dataset is publicly available at \href{https://huggingface.co/datasets/tomhu/codecl}{HuggingFace/tomhu/codecl}, and the related code is open-sourced at \href{https://github.com/Tomsawyerhu/CodeCL}{GitHub/Tomsawyerhu/CodeCL}.

\noindent \textbf{Conflict of Interest}.
The authors declared that they have no conflict of interest.

\noindent \textbf{Clinical Trial Number}: Not Applicable. 

\bibliographystyle{spbasic}
\bibliography{reference}

\end{document}